\documentclass[sn-chicago,iicol,referee]{sn-jnl}% Default with double column layout

\usepackage{graphicx}%
\usepackage{multirow}%
\usepackage{amsmath,amssymb,amsfonts}%
\usepackage{amsthm}%
\usepackage{mathrsfs}%
\usepackage[title]{appendix}%
\usepackage{xcolor}%
\usepackage{textcomp}%
\usepackage{manyfoot}%
\usepackage{booktabs}%
\usepackage{algorithm}%
\usepackage{algorithmicx}%
\usepackage{algpseudocode}%
\usepackage{listings}%
\usepackage{bm}%
\usepackage{float}
\usepackage{placeins}

\begin{document}

% \begin{frontmatter}

\title[TOBART-1]{Type I Tobit Bayesian Additive Regression Trees for Censored Outcome Regression}


\author*[1]{\fnm{Eoghan} \sur{O'Neill}}\email{oneill@ese.eur.nl}
% ORCID ID: 0000-0002-1274-4248



\affil*[1]{ \orgdiv{Econometric Institute}, \orgname{Erasmus University Rotterdam} %\orgaddress{\street{Burg. Oudlaan 50}, \city{Rotterdam}, \postcode{3062 PA}, \state{Zuid-Holland}, \country{The Netherlands}}
}

% \affil[2]{\orgdiv{Department}, \orgname{Organization}, \orgaddress{\street{Street}, \city{City}, \postcode{10587}, \state{State}, \country{Country}}}

% \affil[3]{\orgdiv{Department}, \orgname{Organization}, \orgaddress{\street{Street}, \city{City}, \postcode{610101}, \state{State}, \country{Country}}}


\abstract{Censoring occurs when an outcome is unobserved beyond some threshold value. Methods that do not account for censoring produce biased predictions of the unobserved outcome. This paper introduces Type I Tobit Bayesian Additive Regression Tree (TOBART-1) models for censored outcomes. Simulation results and real data applications demonstrate that TOBART-1 produces accurate predictions of censored outcomes. TOBART-1 provides posterior intervals for the conditional expectation and other quantities of interest. The error term distribution can have a large impact on the expectation of the censored outcome. Therefore, the error is flexibly modeled as a Dirichlet process mixture of normal distributions. An R package is available at \url{https://github.com/EoghanONeill/TobitBART}.}

\keywords{BART, censored regression, regression trees, Bayesian nonparametrics, machine learning.}



\maketitle

\section{Introduction}
\label{sec:intro}

% Introduce task

%The Type I Tobit Model \citep{tobin1958estimation} was implemented in a Bayesian framework by \cite{chib1992bayes}. 

Censoring occurs when, beyond some threshold value, the observed outcome is equal to the threshold instead of the true latent outcome value. For example, scientific equipment can often only make accurate measurements within a known range of outcome values, and observations outside this range are set to its limits. Often the estimand of interest is the conditional expectation or conditional average treatment effect on the outcome before censoring. Estimation of a standard regression model using data without censored values, or with censored observations set equal to threshold values, results in biased estimates. Tobit models directly model the latent outcome and censoring process \citep{tobin1958estimation}.

In this paper, we combine the Bayesian Type I Tobit model \citep{chib1992bayes} with Bayesian Additive Regression Trees \citep{chipman2010bart}. The latent outcome (before censoring) is modeled as a sum-of-trees, which allows for nonlinear functions of covariates. The error term is modeled as a Dirichlet process mixture of normal distributions, as in fully nonparametric BART \citep{george2019fully}. Smooth data generating processes with sparsity are modelled by soft trees with a Dirichlet prior on splitting variable probabilities, as introduced by \cite{linero2018bayesianB}. %This provides a flexible framework for censored outcome prediction.

In simulations and applications to real data, TOBART-1 outperforms a Tobit gradient boosted tree method, Grabit \citep{sigrist2019grabit}, a Tobit Gaussian Process model \citep{groot2012gaussian}, standard linear Tobit, and simple hurdle models based on standard machine learning methods. %\footnote{\cite{deshmukh2020gradient} also describes Gradient Boosted Tree methods for Tobit and Tweedie regression. In the special case of data censored from below at zero, a competing method is Tweedie regression, in particular Tweedie loss based gradient boosting \citep{zhou2020tweedie}. This may or may not be a preferred alternative depending on the distribution of the outcome variable. Furthermore TOBART is more general in that it allows for other limits, two-sided censoring, and nonparametric modelling of the error.} 
Unlike other methods, TOBART-1 accounts for model uncertainty and can non-parametrically model the error term. Posterior intervals are available for censored outcomes, uncensored outcomes, conditional expectations, and probabilities of censoring. Grabit, Gaussian Processes, and other methods rely on cross-validation for parameter tuning and are sensitive to the tuned variance of the error term, whereas TOBART-1 performs well without parameter tuning and accounts for uncertainty in the variance of the error term. %This is consistent with findings for uncensored outcome modelling with standard BART \citep{chipman2010bart}.

%An additional advantage of TOBART over other nonparametric methods such as Gaussian Processes is the ability to model data generating processes with unknown interactions and high-dimensionality.

TOBART-1 with a Dirichlet process mixture of normal distributions for the error term (TOBART-1-NP) removes the restrictive normality assumption often imposed in censored outcome models. We observe that this can lead to more accurate outcome predictions in simulations with non-normally distributed errors, and in real data applications, which may involve non-normally distributed outcomes.\footnote{A Dirichlet process mixture for the error term distribution has previously been included in a censored outcome model by \cite{kottas2009bayesian}.}

A variety of methods have been proposed for nonparametric and semiparametric censored outcome models. \cite{lewbel2002nonparametric} describe a local linear kernel estimator for the setting in which both the uncensored outcome mean function of regressors and error distribution are unknown. \cite{fan1994censored} describe a quantile-based local linear approximation method. \cite{huang2021novel} introduces a semiparametric method involving B-splines. \cite{chen2005nonparametric} use a local polynomial method. Other papers on the topic of semiparametric and nonparametric censored outcome regression include \cite{cheng2021semiparametric, heuchenne2007location, heuchenne2010estimation, huang2019estimation}; and \cite{oganisian2021bayesian}. Gaussian Process censored outcome regression methods are applied by \cite{groot2012gaussian, cao2018model,gammelli2020estimating, gammelli2022generalized} and \cite{basson2023variational}. \cite{zhang2021deep} and \cite{wu2018deep} implement censored outcome neural network methods. %\cite{taddy2010bayesian} describe Nonparametric Bayesian Tobit quantile regression using multivariate normal mixtures.

A number of recent papers have considered Tobit model selection and regularization. \cite{zhang2012focused} describe Focused Information Criteria based Tobit model selection and averaging. \cite{jacobson2022high} provide theoretical and empirical results for Tobit with a Lasso penalty and a folded concave penalty (SCAD). \cite{muller2016censored} and \cite{soret2018lasso} describe a LASSO penalized censored outcome models. %based on \cite{powell1984least}.
\cite{bradic2016robust} study robust penalized estimators for censored outcome regression. %Other papers that describe penalized linear censored outcome methods include \cite{soret2018lasso}.

The Bayesian Tobit literature includes quantile regression methods \citep{ji2012model, yu2007bayesian, alhamzawi2016bayesian}, and Bayesian elastic net Tobit \citep{alhamzawi2020new}. \cite{ji2012model} account for model uncertainty by implementing Tobit quantile regression with Stochastic Search Variable Selection. However, the outcome and latent variable are modeled as linear functions of covariates. TOBART-1 provides a competing approach to the methods referenced above that does not impose linearity.

The remainder of the paper is structured as follows: In section \ref{methods_sec} we describe the TOBART-1 model and Markov chain Monte Carlo (MCMC) implementation, section \ref{simulation_sec} contains simulation studies for prediction and treatment effect estimation with censored data, section \ref{application_sec} contains applications to real world data, and section \ref{conclusion_sec} concludes the paper.

\section{Methods}\label{methods_sec}

\subsection{Review of Bayesian Additive Regression Trees (BART)}\label{bart_review}

% \noindent \textbf{Description of Model and Priors}
%copy and paste from existing documents

% \medskip

Suppose there are $n$ observations, and the $n \times p$ matrix of explanatory variables, $X$, has $i^{th}$ row $x_i=[x_{i1},...,x_{ip}]$. Following the notation of \cite{chipman2010bart}, let $T$ be a binary tree consisting of a set of interior node decision rules and a set of terminal
nodes, and let $M = \{ \mu_1 , ..., \mu_b \}$ denote a set of parameter values associated with each of the $b$ terminal nodes of $T$. The interior node decision rules are binary splits of the predictor space into the sets $\{ x_{is} \le c \}$ and $\{ x_{is} > c \}$ for continuous $x_{s}$. Each observation's $x_i$ vector is associated with a single terminal node of $T$, and is assigned the $\mu$ value associated with this terminal node. For a given $T$ and $M$, the function $g(x_i;T,M)$ assigns a $\mu \in M$
to $x_i$. %This gives the single tree model $Y \sim g(x_i;T,M) + \varepsilon \ , \ \varepsilon \sim N(0,\sigma^2)$  \citep{chipman1998bayesian}.

For the standard BART model, the outcome is determined by a sum of trees,
$$Y_i = \sum_{j=1}^m g(x_i ; T_j, M_j)+\varepsilon_i$$
where $g(x_i;T_j,M_j)$ is the output of a decision tree. $T_j$ refers to a decision tree indexed by $j=1,...,m$, where $m$ is the total number of trees in the model. $M_j$ is the set of terminal node parameters of $T_j$, and $\varepsilon_i \overset{i.i.d}{\sim} N(0, \sigma^2)$.

Prior independence is assumed across trees $T_j$ and across terminal node means $M_j = (\mu_{1j}...\mu_{b_j j})$ (where $1,...,b_j$ indexes the terminal nodes of tree $j$). The form of the prior used by \cite{chipman2010bart} is
$p(M_1,...,M_m,T_1,...,T_m,\sigma) \propto \left[ \prod_j \left[ \prod_k p(\mu_{kj}|T_j) \right] p(T_j)\right]p(\sigma) $
%
%For BCF \citep{hahn2020bayesian}, we are interested in the model $f(x_i,z_i)= \mu(x_i,\hat{\pi}_i)+\tau(x_i)z_i$, where $\mu(x_i,\hat{\pi}_i)$ and $\tau(x_i) $ are separate sum of tree models. Let $T_{\mu j}$ and $T_{\tau j}$ denote  trees in $\mu(x_i,\hat{\pi}_i)$ and $\tau(x_i) $ respectively, and let $M_{\mu j}$ and $M_{\tau j}$ denote the terminal node parameters for $T_{\mu j}$ and $T_{\tau j}$ respectively.  The BCF prior can be written as:
%
%$$p(M_{\mu 1},...,M_{\mu m_{\mu}},T_{\mu 1},...,T_{\mu m_{\mu}}, M_{\tau 1},...,M_{\tau m_{\tau}}, T_{\tau 1},...,T_{\tau m_{\tau}}, \sigma) $$
%$$\propto \left[ \prod_j \prod_i p(\mu_{ij}|T_{\mu j})p(T_{\mu j})\right] \left[ \prod_j \prod_i p(\tau_{ij}|T_{\tau j})p(T_{\tau j})\right] p(\sigma)   $$
%
where $\mu_{kj} | T_j \overset{i.i.d}{\sim} N(0,\sigma_{\mu}^2)$ where $\sigma_{\mu} = \frac{0.5}{\kappa \sqrt{m}}$ and $\kappa$ is a user-specified hyper-parameter. %In BCF, \cite{hahn2020bayesian} set a half-Cauchy prior on the scale of $\mu_{ij}$ with prior median equal to twice the marginal standard deviation of $Y$, and set a half-Normal prior on the scale of $\tau_{ij} | T_j $.

\cite{chipman2010bart} set a regularization prior on the tree size and shape $p(T_j)$. The probability that a given node within a tree $T_j$ is split into two child nodes is $\alpha (1+d_h)^{-\beta}$, where $d_h$ is the depth of (internal) node $h$, and the parameters $\alpha$ and $\beta$ determine the size and shape of $T_j$ respectively. \cite{chipman2010bart} use uniform priors on available splitting variables and splitting points. %Thus $p(T_j) = \prod_{h=1}^{b_j-1} \alpha (1+d_h)^{-\beta}  \prod_{k=1}^{b_j} (1 - \alpha (1+d_k)^{-\beta} ) $, where $h$ indexes the internal nodes of the tree $T_j$, and $k$ indexes the terminal nodes. %In BCF, \cite{hahn2020bayesian} use: for $\mu(x)$, the default suggestions of 200 trees, $\beta=2, \alpha=0.95$, and for $\tau(x)$, there is stronger regularization, with 50 trees, $\beta=3, \alpha=0.25$.
The model precision $\sigma^{-2}$ has a conjugate prior distribution $\sigma^{-2} \sim Ga(\frac{v}{2}, \frac{v \lambda}{2})$ with degrees of freedom $v$ and scale $\lambda$. %The same prior is used for the model precision in BCF.

Samples from $p((T_1, M_1),...,(T_m,M_m), \sigma | y)$ can be made by a Bayesian backfitting MCMC algorithm. This algorithm involves $m$ successive draws from 
$(T_j , M_j )| T_{(j)} , M_{(j)} , \sigma , y $ for $j=1,...,m$, where $T_{(j)} , M_{(j)} $ are the trees and parameters for all trees except the $j^{th}$ tree, followed by a draw of $\sigma $ from the full conditional $\sigma | T_1,...,T_m,M_1,...,M_m,y$. 
% A set of draws induces the sum of trees function $f^*(.)= \sum_{j=1}^m g(. \ ; T_j^*, M_j^*)$. 
After burn-in, the sequence of $f^*$ draws, $f_1^*,...,f_Q^*$, where $f^*(.)= \sum_{j=1}^m g(. \ ; T_j^*, M_j^*)$, is an approximate sample of size $Q$ from $p(f|y)$.
%
% To estimate the unknown function $f(x)$,\footnote{$Y_i = \sum_{j=1}^m f(x_i) +\varepsilon_i \approx \sum_{j=1}^m g(x_i ; T_j, M_j)+\varepsilon_i$} %or predict $Y$ in-sample or out-of-sample, 
% a natural choice is
% $\frac{1}{Q} \sum_{q=1}^Q f_{q}^* (x)$, 
% which approximates $E(f(x)|y)$. 
%Prediction intervals can be obtained from quantiles of the draws $f_{q}^* (x)$. 

%A number of papers describe faster BART implementation algorithms and improved sampling methods, including parallelized BART \citep{pratola2014parallel}, particle Gibbs algorithms \citep{lakshminarayanan2015particle}% (an extension of single tree Sequential Monte Carlo ideas proposed by \citep{lakshminarayanan2013top}), more efficient Metropolis-Hastings proposals \citep{pratola2016efficient}, Consensus Monte Carlo \citep{scott2016bayes}, a likelihood-inflated sampling algorithm \citep{entezari2018likelihood}, and Accelerated BART (X-BART), which uses a stochastic hill climbing algorithm as a greedy stochastic approximation to MCMC \citep{he2018accelerated}.
%

\subsection{Soft Trees and Sparse Splitting Rules}

In addition to the standard Bayesian tree model for $f(\bm{x}_i)$ described in section \ref{bart_review}, we also implement TOBART and TOBART-NP with soft trees and sparse splitting rules as described by \cite{linero2018bayesianB}. Predictions from soft trees are weighted linear combinations of all terminal node parameter values, with the weights being  functions of distances between covariates and splitting points. The prediction from a single tree function is 
$$ g(\bm{x}_i ; T_j, M_j)  = \sum_{\ell = 1}^{L_j} \mu_{j,\ell}  \xi (\bm{x}_i, T_j, \ell) $$
$$\xi (\bm{x}_i, T_j, \ell) = \prod_{b \in \mathcal{A}(\ell)}  \zeta \left( \frac{ x_{{j_b}} - C_b}{\tau_b} \right)^{ \mathbb{I} \{  x_{j_b} > C_b  \} } \times $$
$$\Big\{ 1 - \zeta \left( \frac{ x_{{j_b}} - C_b}{\tau_b} \right) \Big\}^{ \mathbb{I} \{  x_{j_b} \le C_b  \} }  $$
% $$  = \prod_{b \in \mathcal{A}(\ell)}  \exp \left(  \frac{ x_{{j_b}} - C_b}{\tau_b}  \right)^{ \mathbb{I} \{  x_{j_b} > C_b  \}} \Bigg/ \Bigg(1 + \exp \left(  \frac{ x_{{j_b}} - C_b}{\tau_b}  \right) \Bigg)   $$
%
where $L_j$ is the number of leaves in the $j^{th}$ tree, $\mu_{j,\ell}$ is the $\ell^{th}$ terminal node parameter of the $j^{th}$ tree, $\mathcal{A}(\ell)$ denotes the set of ancestor nodes of terminal node $\ell$. The splitting variable, splitting point, and bandwidth parameter at internal node $b$ are denoted by $x_{{j_b}}$, $C_b$, and $\tau_b$ respectively. The gating function $\zeta $ is the logistic function $ \zeta(x) = (1+\exp(-x))^{-1}$.

Sparse splitting rules are introduced by placing a Dirichlet prior on the splitting probabilities $(s_1,\dots, s_p) \sim \mathcal{D} (\frac{a}{p},\dots, \frac{a}{p})$. The parameter $a$ controls the level of sparsity and has the prior $\text{Beta}(0.5,1)$. \cite{linero2018bayesianB} demonstrate that soft trees allow BART to model smooth functions, and the Dirichlet prior on splitting probabilities adapts to unknown levels of sparsity to provide improved predictions on high dimensional data sets.

\subsection{Type I Tobit and TOBART}
\subsubsection{Type I Tobit Model}
The Type I Tobit model with censoring from below at $a$ and censoring from above at $b$ is:
$$ Y_i^* = \bm{x}_i \bm{\beta} + \varepsilon_i \ , \ \varepsilon_i \sim i.i.d. \ N(0, \sigma^2) $$
$$ Y_i = \begin{cases}
	a \ \text{if} \ Y_i^* \le a \\
	Y_i^* \ \text{if} \ a < Y_i^* < b \\
	b \ \text{if} \ b \le Y_i^*  
\end{cases} $$
where a normal prior is placed on $\beta$, and an inverse gamma prior is placed on $\sigma^2$ \citep{chib1992bayes}.

\subsubsection{Type I TOBART Model}

The Type I TOBART model replaces the linear combination $\bm{x}_i \bm{\beta}$ with the sum-of-trees function $f(\bm{x}_i)$ :
$$ Y_i^* = f(\bm{x}_i)  + \varepsilon_i \ , \ \varepsilon_i \sim i.i.d. \ N(0, \sigma^2) $$
$$ Y_i = \begin{cases}
	a \ \text{if} \ Y_i^* \le a \\
	Y_i^* \ \text{if} \ a < Y_i^* < b \\
	b \ \text{if} \ b \le Y_i^*  
\end{cases} $$
where a BART prior is placed on $ f(\bm{x}_i)$ and an inverse gamma prior is placed on $\sigma^2$. \footnote{$\sigma^{-2} \sim Ga(\frac{v}{2}, \frac{v \lambda}{2})$. For standard BART, $\lambda$ is set such that the $q^{th}$ quantile of the prior distribution of $\sigma$ is the sample standard deviation of the residuals from a linear model. For censored outcomes, this may give poor calibration of the $\sigma$ prior. We consider four options in a simulation study in appendix \ref{tbarts_sims_app}. A sample standard deviation estimate from an intercept-only Tobit model generally gives good results, although often there is little difference across $\lambda$ values.}

% \subsubsection{Type I Tobit Gibbs Sampler}

% Tobit can be implemented by MCMC with data augmentation \citep{chib1992bayes}. The variable $y_i^*$ is observed for uncensored outcomes, and is sampled from its full conditional for censored outcomes.

% The $y_i^*$ values for censored observations are sampled from the full conditionals:
% %
% $$ y_i^* \sim \begin{cases}
% 	\mathcal{N}(\bm{x}_i' \bm{\beta}, \sigma^2)	I(-\infty,a) \ \text{if} \ y_i^* < a \\
% 	\mathcal{N}(\bm{x}_i' \bm{\beta}, \sigma^2) 	I(b, \infty)\ \text{if} \ y_i^* > b
% \end{cases} $$
% %
% The full conditionals for $\bm{\beta}$ and $\sigma^2$ are standard full conditionals for Bayesian linear regression of $y_i^*$ on $\bm{x}_i$

\subsubsection{Type I TOBART Gibbs Sampler}
Tobit can be implemented by MCMC with data augmentation \citep{chib1992bayes}. The realization, $y_i^*$, of the variable $Y_i^*$ is observed for uncensored outcomes, and is sampled from its full conditional for censored outcomes.
%
% The $y_i^*$ values for censored observations are sampled from the full conditionals:
%
$$y_i^* = y_i \text{ if } y_i \in (a,b) \ , \  \text{ and } $$ 
$$y_i^* \sim \begin{cases}
	\mathcal{TN}_{[-\infty,a]}(f(\bm{x}_i) , \sigma^2)	 \ \text{if} \ y_i = a \\
	\mathcal{TN}_{[b,\infty]}(f(\bm{x}_i) , \sigma^2) 	\ \text{if} \ y_i = b
\end{cases} $$
where $\mathcal{TN}_{[l,u]}$ denotes a normal distribution truncated to the interval $[l,u]$.  
The full conditionals for $f(\bm{x}_i)$ and $\sigma^2$ are standard full conditionals for BART with $y_i^*$ as the dependent variable and $\bm{x}_i$ as the potential splitting variables. 
Appendix \ref{mcmc_app} contains a description of a sampler that produces draws $f^{(1)}(\bm{x}_i), \hdots,f^{(D)}(\bm{x}_i)$ and $\sigma^{(1)},\hdots,\sigma^{(D)}$.

\subsubsection{Predicting Outcomes with TOBART}\label{outcomepred_sibsec}

The conditional mean of the latent variable is $f(\bm{x}_i)$. If censoring is also applied to the test data, then the outcomes are predicted by averaging the standard Tobit expectation formula across MCMC iterations:
%using the conditional expectation calculated as described below. For each MCMC iteration, the standard Tobit expectation formula can be applied:

For all MCMC iterations $d=1,...,D$ calculate
$$ E[Y_i|X_i=\bm{x}_i, f^{(d)}, \sigma^{(d)}] = a \Phi \Big(\frac{a - f^{(d)}(\bm{x}_i)}{\sigma^{(d)}}  \Big)  + $$
$$ f^{(d)}(\bm{x}_i) \Bigg[ \Phi \Big(\frac{b - f^{(d)}(\bm{x}_i)}{\sigma^{(d)}}  \Big) - \Phi \Big(\frac{a - f^{(d)}(\bm{x}_i)}{\sigma^{(d)}}  \Big)  \Bigg] $$
$$ + \sigma^{(d)} \Bigg( \phi \Big(\frac{a - f^{(d)}(\bm{x}_i)}{\sigma^{(d)}}  \Big) - \phi \Big(\frac{b - f^{(d)}(\bm{x}_i)}{\sigma^{(d)}}  \Big) \Bigg) + $$
$$ b \Bigg[ 1 - \Phi \Big(\frac{b - f^{(d)}(\bm{x}_i)}{\sigma^{(d)}}  \Big) \Bigg]$$
The predicted outcome is $\frac{1}{D} \sum_{d=1}^D  E[Y_i|X_i=\bm{x}_i, f^{(d)}, \sigma^{(d)}]$. 
The expectation conditional on the outcome not being in the censored range is:
$$E[Y_i| a < Y_i < b, X_i=\bm{x}_i, f^{(d)}, \sigma^{(d)}] =  $$
$$ f^{(d)}(\bm{x}_i)   + \sigma^{(d)} \frac{ \phi \Big(\frac{a - f^{(d)}(\bm{x}_i)}{\sigma^{(d)}}  \Big) - \phi \Big(\frac{b - f^{(d)}(\bm{x}_i)}{\sigma^{(d)}}  \Big) }{  \Phi \Big(\frac{b - f^{(d)}(\bm{x}_i)}{\sigma^{(d)}}  \Big) - \Phi \Big(\frac{a - f^{(d)}(\bm{x}_i)}{\sigma^{(d)}}  \Big) }  $$

\subsection{Nonparametric Type I TOBART}

\subsubsection{Nonparametric Type I TOBART Model}

The accuracy of the conditional expectation of the TOBART model depends on the validity of the assumption of normality of the errors. More general censored outcomes can be modelled by assuming a Dirichlet Process mixture distribution for the error terms.
$$ y_i^* = f(\bm{x}_i)  + \varepsilon_i \ ,\ y_i = \begin{cases}
	a \ \text{if} \ y_i^* \le a \\
	y_i^* \ \text{if} \ a < y_i^* < b \\
	b \ \text{if} \ b \le y_i^*  
\end{cases} $$
$$\varepsilon_i \sim i.i.d. \ N(\gamma_i, \sigma_i^2) \ , \  \vartheta_i = (\gamma_i, \sigma_i) \sim G $$ 
$$ G \sim \mathcal{DP}(G_0, \alpha)$$
%
% $$ \vartheta_i = (\gamma_i, \sigma_i) \sim G \ , \ G \sim \mathcal{DP}(G_0, \alpha)$$
%
% $$ G \sim \mathcal{DP}(G_0, \alpha) $$
%
The distribution of the error term is specified similarly to \cite{george2019fully}. The base distribution $G_0$ is defined as follows:
$$ p(\gamma, \sigma | \nu, \lambda_1, \gamma_0, k_0) = p(\sigma | \nu, \lambda) p(\gamma | \sigma, \gamma_0, k_0) $$
$$ \sigma^2 \sim \frac{\nu \lambda}{\chi_{\nu}^2} \ , \ \gamma | \sigma \sim \mathcal{N} \Big(\gamma_0, \frac{\sigma^2}{k_0} \Big) $$
where, in contrast to the standard BART prior of \cite{chipman2010bart}, $\nu$ is set to $10$ instead of $3$.\footnote{\cite{george2019fully} recommend $\nu=10$ as the spread of the error increases when there are many components and the spread of a single components can be reduced by increasing $\nu$. This gives better results than $\nu=3$ for some DGPs in a simulation study in Appendix \ref{tbarts_sims_app}.} The parameter $\lambda$ is set such that the $q^{th}$ quantile of the prior distribution of $\sigma$ is the sample standard deviation of the outcome, or of the residuals from a linear model. For TOBART-NP, $q=0.9$ instead of $0.95$. \footnote{This is complicated by the censoring of the outcome. Some options are:
	1. Estimate the standard deviation assuming that censored outcome is normally distributed. 
	2. Estimate the standard deviation of a linear type I Tobit model (contains option 1 as a special case but not feasible when there are more regressors than observations). 
	3. Estimate the standard deviation of the censored outcome without accounting for censoring. We use option 2 for TOBART-NP.
} 
The prior on $\alpha$ is the $\alpha \sim \Gamma(2,2)$ prior introduced by \cite{escobar1995bayesian} and applied by \cite{van2011bayesian}.\footnote{The TobitBART package also includes an option for the prior described by \cite{rossi2014bayesian} and \cite{george2019fully}, $p(\alpha) \propto \left( 1 - \frac{\alpha - \alpha_{min}}{\alpha_{max} - \alpha_{min}} \right)^{\psi}$, where $\alpha_{min}$ and $\alpha_{max}$ are set so that the modal numbers of components are $I_{min} = 1$ and $I_{max} = [(0.1)n]$ respectively, and $\psi = 0.5$.}

The outcome is scaled by subtracting the sample mean before applying the Gibbs sampler, therefore \cite{george2019fully} set $\gamma_0 = 0$.\footnote{However, the mean cannot be estimated for censored data without making further assumptions. Options include: 1. Estimate the mean (and variance) of a censored normal distribution. 2. Calculate the sample mean of the censored outcome without accounting for censoring. We use option 1.%, which can be implemented by estimating a Tobit regression with intercept only.
} The parameter $k_0$ is scaled with the marginal distribution of $\gamma$ ( $\gamma \sim \frac{\sqrt{\lambda } }{ \sqrt{k_0 } } t_{\nu} $). Given $k_s$ (set to $10$ by default), $k_0$ is set such that $ \max_{i=1,...,n} |e_i| = k_s \frac{\sqrt{\lambda } }{ \sqrt{k_0 } } $ 
where $k_s = 10$. and $e_1,...,e_n$ are the residuals from a linear model.\footnote{The residuals likely underestimate the true errors for censored observations.} 
% 
%
% \todo[inline]{The prior on $\sigma$ can also be written as $\sigma^2 \sim \mathcal{IG}\Big( \frac{\nu}{2}, \frac{\nu \lambda}{2}  \Big)$. }
%
%
The Gibbs sampler for TOBART-NP is described in Appendix \ref{mcmc_app}.

For each MCMC iteration, $d$, and observation $i$, we obtain $\vartheta_i^{(d)} = (\gamma_i^{(d)}, \sigma_i^{(d)})$. The conditional expectation, $E[y_i|\bm{x}_i, f^{(d)}, \gamma_i^{(d)}, \sigma^{(d)}] $,  is calculated as outlined in section \ref{outcomepred_sibsec}.

\subsection{Treatment Effect Estimation for Censored Outcomes}

Let a binary variable $T_i$ equal $1$ if unit $i$ is assigned to treatment and $0$ if $i$ is assigned to the control group. The potential outcomes under treatment and control group allocation are denoted by $Y_i(1)$ and $Y_i(0)$ respectively. Similarly, the potential outcomes of the latent outcome are denoted by $Y_i^*(1), Y_i^*(0)$. Assume the data generating process is as follows:
$$ Y_i^* = \mu(\bm{x}_i) + \tau(\bm{x}_i) T_i + \varepsilon_i \ , \  \varepsilon_i  \sim \mathcal{N}(0,\sigma^2) $$
$$ Y_i = \begin{cases}
	a \ \text{if} \ Y_i^* \le a \\
	Y_i^* \ \text{if} \ a < Y_i^* < b \\
	b \ \text{if} \ b \le Y_i^* 
\end{cases}
$$
where $\mu(\bm{x}_i)$ and $\tau(\bm{x}_i)$ are possibly nonlinear functions of covariates. Assume conditional unconfoundedness, i.e. $Y_i^*(1),Y_i^*(0) \perp T_i | X_i$ . The estimand is the conditional average treatment effect on $Y_i^*$, i.e., $ E[Y_i^*(1) - Y_i^*(0) | X_i = \bm{x}_i] = \tau(\bm{x}_i)  $. 
However, a model naively trained on only uncensored outcomes estimates the following effects \footnote{This bias occurs if all the uncensored observations are included in one regression and differences in predictions for $T_i=1$ and $T_i=0$ are obtained, i.e. an S-learner approach \citep{kunzel2019metalearners}, or if the two conditional expectations are obtained from separate regressions for treated and untreated uncensored observations, i.e. a T-Learner approach. In both cases, the conditional expectations are not equal to the  expectation of the latent outcome.}
$$ E[Y_i(1) | a < y_i < b, X_i = \bm{x}_i] - $$
$$ E[Y_i(0) | a < y_i < b, X_i = \bm{x}_i] =  \tau(\bm{x}_i)   + $$
$$ \sigma  \Bigg( \frac{ \phi \Big(\frac{a - ( \mu(\bm{x}_i) + \tau(\bm{x}_i)))}{\sigma}  \Big) - \phi \Big(\frac{b - ( \mu(\bm{x}_i) + \tau(\bm{x}_i))  }{\sigma}  \Big) }{  \Phi \Big(\frac{b - ( \mu(\bm{x}_i) + \tau(\bm{x}_i))  }{\sigma}  \Big) - \Phi \Big(\frac{a - ( \mu(\bm{x}_i) + \tau(\bm{x}_i))      }{\sigma}  \Big) }  - 
$$
$$\frac{ \phi \Big(\frac{a - \mu(\bm{x}_i)}{\sigma}  \Big) - \phi \Big(\frac{b - \mu(\bm{x}_i)}{\sigma}  \Big) }{  \Phi \Big(\frac{b - \mu(\bm{x}_i)}{\sigma}  \Big) - \Phi \Big(\frac{a - \mu(\bm{x}_i)}{\sigma}  \Big) }
\Bigg) \ . $$
% where the second term %follows from assumptions on the error distribution, and can be modelled by 
% can accurately be included in the estimated conditional expectations from a sufficiently flexible nonparametric method without restrictive assumptions on the error term.
A sufficiently flexible nonparametric method, without restrictive assumptions on the error term, will produce estimates that approximate the expression above. 
A model naively trained on the full data set with censoring similarly gives biased estimates (see Appendix \ref{TEbias_app}). 
By directly modelling $Y_i^*$, censored outcome models avoid the bias described above. Similar biases occur if the error term is not normally distributed.

% \subsubsection{Nonparametric Type I Tobit Gibbs Sampler}

\section{Simulation Studies}\label{simulation_sec}

\subsection{Description of Prediction Simulations}

We adapt the data generating processes (DGPs) introduced by \cite{friedman1991multivariate} to a censored regression setting. This DGP has often been applied in comparisons of semiparametric regression methods. We also make use of the censored outcome simulations described by \cite{groot2012gaussian}, \cite{sigrist2019grabit}, and  \cite{jacobson2022high} for fair comparison against competing methods with existing synthetic censored data. %We consider DGPs with one-sided and two-sided censoring.

The covariates $x_1,....,x_p$ are independently sampled from the uniform distribution on the unit interval. The outcome before censoring is generating from one of the following functions:
\begin{itemize}
	\item $ y^* =  10 \sin (\pi x_1 x_2) + 20 (x_3 - 0.5)^2 + 10 x_4 + 5 x_5 + \varepsilon \ , \  \varepsilon \sim \mathcal{N}(0,\sigma^2) $ with censoring from below at the $15^{th}$ percentile of the training data $y^*$ values \citep{friedman1991multivariate}.\footnote{The original Friedman simulations did not involve censoring.}
	\item $ y^* =  10 \sin (\pi x_1 x_2) + 20 (x_3 - 0.5)^2 + 10 x_4 + 5 x_5 + \varepsilon \ , \  \varepsilon \sim \mathcal{N}(0,\sigma^2) $ with censoring from below at the $15^{th}$ percentile of the training data $y^*$ values, and from above at the $85^{th}$ percentile of the training data $y^*$ values \citep{friedman1991multivariate}.
	\item $ y^* =  6 (x_1 - 2)^2 \sin (2(6x_1 - 2) ) + \varepsilon \ , \  \varepsilon \sim \mathcal{N}(0,\sigma^2) $ with censoring from below at the $40^{th}$ percentile of the training data $y^*$ values \citep{groot2012gaussian}.
	\item $ y^* =  \sum_{k=1}^5 0.3 \max(x_k,0) + \sum_{k=1}^3 \sum_{j=k+1}^4 \max (x_k x_j,0) + \varepsilon \ , \  \varepsilon \sim \mathcal{N}(0,\sigma^2) $ with censoring from above at the $95^{th}$ percentile of the training data $y^*$ values \citep{sigrist2019grabit}. For this simulation, $x_1,....,x_p$ are uniformly distributed on $[-1,1]$ instead of $[0,1]$.\footnote{This simulation differs somewhat from the original simulation of \cite{sigrist2019grabit} for which the variable determining censoring was not perfectly correlated with the observed outcome before censoring.}
	\item $ y^* =  3 + 5 x_1 + x_2 + \frac{x_3}{2} - 2 x_4 + \frac{x_5}{10} + \varepsilon \ , \  \varepsilon \sim \mathcal{N}(0,\sigma^2) $ with censoring from below at the $25^{th}$ percentile of the training data $y^*$ values \citep{jacobson2022high}.
\end{itemize}
The variance of the error, $\sigma^2$, is set to 1. See the Supplementary Appendix (Online Resource 1) for the results obtained from simulations with $\sigma \in \{0.1, 2\}$. %, and additional results obtained with a mixture of normal distributions for the error. 
%
%We consider a range of values for the standard deviation of the error term: $\sigma \in \{0.1, 1, 2\}$. 
We also consider deviations from the assumption of normally distributed errors. In particular, we include results for simulations in which $\varepsilon$ is generated from Skew-t, and $\text{Weibull}(1/2, 1/5)$ distributions.\footnote{\cite{bradic2016robust} considered $\text{Weibull}(1/2, 1/5)$ errors in a simulation study.} The number of covariates, $p$, is set to $30$. We generate 500 training and 500 test observations. %Then, for the one-sided simulation the outcome is censored from below at the $15^{th}$ percentile of the training data $y^*$ values. For the two-sided simulation, the outcome is also censored from above at the $85^{th}$ percentile of the training data $y^*$ values.

\subsection{Prediction Simulation Results}\label{sim_results}

We compare the performance of TOBART-1, TOBART-1-NP, Soft TOBART-1, and Soft TOBART-1-NP against Grabit \citep{sigrist2019grabit}, linear Tobit \citep{tobin1958estimation}, BART \citep{chipman2010bart}, Random Forests (RF) \citep{breiman2001random}, Gaussian Processes, and a Tobit Gaussian Process model \citep{groot2012gaussian}.\footnote{Standard BART for continuous outcomes is trained on censored outcomes. Probit BART is trained on a binary variable indicating censorship. Similarly, Random Forests are separately trained on continuous censored outcomes and a binary censorship indicator.} The results for a Gaussian Process (GP) with only 5 variables (always including all informative variables) are included because GPs were observed to produce inaccurate predictions when applied to data with 30 variables.\footnote{The GP Matlab code was obtained from \url{https://www.cs.ru.nl/~perry/software/tobit1.html}.} Censored outcome predictions are evaluated using Mean Squared Error (MSE), and predicted probabilities of censoring are evaluated using the Brier Score.\footnote{See the Supplementary Appendix (Online Resource 1) for implementation details and parameter settings.} \footnote{Latent outcome predictions similarly demonstrate that TOBART outperforms other methods, and these results are available on request. However, it is unsurprising that Tobit based latent outcome predictions outperform naive approaches due to the aforementioned censoring bias. } All results are averaged over 5 repetitions. \footnote{Computational times are included in Appendix \ref{comp_time_app}.}

The results for simulations with normally distributed errors are presented in Tables \ref{allsim_mse_tab} and \ref{allsim_brier_tab}. The TOBART algorithms generally outperform competing methods across all DGPs, except unsurprisingly for the linear \cite{jacobson2022high} simulations linear Tobit is outperformed only by Soft TOBART. TOBART-NP can slightly improve on TOBART in some cases, but generally the results are similar when errors are normally distributed. The differences in criteria across methods are small for the more linear DGPs from \cite{sigrist2019grabit} and \cite{jacobson2022high}, as linear Tobit is designed for a linear DGP, and the nonlinear methods BART and RF can model the relatively simple response surface well. It is worth noting that TOBART outperforms Grabit even though the true standard deviation, $\sigma=1$, is included as one of five possible Grabit hyperparameter values in cross-validation. The same pattern of results can be observed for simulations with $\sigma =0.1$ and $\sigma = 2$ in the Supplementary Appendix. The Supplementary Appendix contains comparisons of Area Under the Curve for all methods and DGPs, from which similar conclusions can be drawn.

The results for Skew-t and Weibull distributed errors are also presented in Tables \ref{allsim_mse_tab} and \ref{allsim_brier_tab}.\footnote{Results for t-distributed errors with $\nu=3$ are in the Supplementary Appendix.} The TOBART models outperform all other methods for almost all DGPs and criteria. % For most DGPs, the differences between the TOBART and TOBART-NP results are relatively small. %The exception is a smaller MSE achieved by standard BART in the \cite{groot2012gaussian} simulations, which contain many censored observations.%, and a very small difference in AUC relative to linear Tobit for the linear DGP from \cite{jacobson2022high}. 
The results for the Weibull distribution generally favour TOBART-NP and Soft TOBART-NP, indicating that there is some improvement from the Dirichlet Process model when the errors are sufficiently non-Gaussian. %Grabit performs best in terms of MSE for the \cite{groot2012gaussian} simulations. Standard BART is the best method in terms of Brier scores for the \cite{groot2012gaussian} and \cite{jacobson2022high} DGPs. For the \cite{groot2012gaussian} DGP, and the relatively simple DGPs of \cite{sigrist2019grabit} and \cite{jacobson2022high}, TOBART-NP can give some improvement relative to standard TOBART. 

%We observe similar results to \cite{george2019fully} in that TOBART-NP and TOBART are likely to give similar predictions of $f(\bm{x}_i)$, but for some DGPs TOBART-NP may give more accurate uncertainty quantification. %Overall, TOBART-NP does not consistently deliver large improvements over standard TOBART. This might be a result of the similarities of the error distributions to the normal distribution, or a failure of the Markov Chain to converge. The numbers of burn-in and post-burn-in draws, 5000 and 10000 respectively, are quite small, particularly considering the difficulty in achieving convergence for linear Tobit observed by \cite{omori2007efficient}.\footnote{However, large numbers of MCMC draws and MCMC diagnostics would be costly in terms of computational and researcher time, and therefore we restrict our simulations to relatively few MCMC iterations for fair comparison across methods and to reflect how these methods are likely to be implemented in practice. \cite{george2019fully} use the same number of MCMC iterations in an example of fully nonparametric BART.}

% Tables \ref{sim_normbeta_latent_cov_tab} to \ref{sim_weibullt_latent_len_tab}  in Appendix \ref{predint_sec_app} display the average coverage and length of $95\%$ prediction intervals for the latent outcomes. Tables \ref{sim_normbeta_obs_cov_tab} to \ref{sim_weibullt_obs_len_tab}   in Appendix \ref{predint_sec_app} display the average coverage and length of $95\%$ prediction intervals for the observed outcomes. 

The average coverage and length of $95\%$ prediction intervals for the latent outcomes and the observed outcomes are given in the Supplementary Appendix (Online Resource 1). For most DGPs and error distributions, TOBART and Soft TOBART provides the closest to 95\% coverage of prediction intervals for both latent and observed outcomes. For some DGPs with non-normal errors, the more conservative intervals produced by TOBART-NP and Soft TOBART-NP provide better coverage.

%\todo[inline]{NOTE THAT THESE ARE OBSERVED OUTCOME MSE VALUES. CENSORED OUTCOME MODELS OUTPERFORM STANDARD METHODS EVEN MORE WHEN PREDICTING LATENT OUTCOMES}

% \subsubsection{Normal Distribution, Standard Deviation 1}

% RUN TWO SIDED FRIEDMAN AGAIN. MISTAKE IN PROBABILITIES?

\begin{table*}[ht]
	\centering
	\begin{tabular}{l|p{1.1cm}p{1.1cm}p{1.5cm}p{1.1cm}p{1.3cm}p{2.2cm}}
		\hline
		Data  & Tobit & BART & RF & Grabit & TOBART & TOBART NP \\ 
		\hline
		\multicolumn{7}{c}{normal distribution, $\sigma=1$} \\
		\hline
		\cite{friedman1991multivariate} & 4.764 & 1.765 & 4.559 & 2.291 & 1.162 & 1.154 \\ 
		\cite{friedman1991multivariate} 1 side &  6.444 & 1.768 & 6.004 & 2.743 & 1.457 & 1.509 \\ 
		\cite{groot2012gaussian} & 12.886 & 2.247 & 4.612 & 0.702 & 0.631 & 0.617 \\ 
		\cite{jacobson2022high}  & 0.694 & 0.722 & 0.855 & 0.739 & 0.718 & 0.720 \\ 
		\cite{sigrist2019grabit}  &  1.353 & 1.142 & 1.170 & 1.146 & 1.072 & 1.074 \\ 
		\hline
		\multicolumn{7}{c}{Skew-t distribution, $location = 1$, $scale = 1$, $\nu=4$}\\
		\hline
		\cite{friedman1991multivariate} & 5.075 & 2.172 & 4.790 & 2.495 & 1.648 & 1.597 \\ 
		\cite{friedman1991multivariate} 1 side & 6.815 & 2.352 & 6.190 & 3.197 & 2.163 & 2.060 \\ 
		\cite{groot2012gaussian} & 14.700 & 2.999 & 5.563 & 1.204 & 1.149 & 1.071 \\ 
		\cite{jacobson2022high} & 1.195 & 1.264 & 1.301 & 1.271 & 1.238 & 1.184 \\ 
		\cite{sigrist2019grabit} & 1.519 & 1.333 & 1.337 & 1.340 & 1.279 & 1.267 \\  
		\hline
		\multicolumn{7}{c}{Weibull distribution, $shape = 0.5$, $scale = 0.2$} \\
		\hline
		\cite{friedman1991multivariate} & 4.599 & 1.502 & 4.576 & 2.099 & 0.871 & 0.811 \\ 
		\cite{friedman1991multivariate} 1 side & 6.221 & 1.624 & 5.962 & 2.650 & 1.344 & 1.154 \\ 
		\cite{groot2012gaussian} & 11.297 & 1.746 & 4.071 & 0.760 & 0.787 & 0.648 \\ 
		\cite{jacobson2022high} & 0.721 & 0.814 & 0.873 & 0.783 & 0.960 & 0.696 \\ 
		\cite{sigrist2019grabit} & 0.739 & 0.473 & 0.536 & 0.447 & 0.426 & 0.362 \\ 
		\hline
		%     		 \multicolumn{7}{c}{t distribution, $\nu = 3$} \\
		% \hline
		%     \cite{friedman1991multivariate} & 6.090 & 3.036 & 5.418 & 4.133 & 2.459 & 2.369 \\ 
		%   \cite{friedman1991multivariate} 1 side & 8.595 & 4.085 & 7.578 & 6.498 & 3.667 & 3.538 \\ 
		%   \cite{groot2012gaussian} & 13.204 & 3.378 & 5.558 & 2.030 & 2.028 & 1.964 \\ 
		%   \cite{jacobson2022high} & 2.328 & 2.488 & 2.489 & 2.461 & 2.394 & 2.306 \\ 
		%   \cite{sigrist2019grabit} & 2.799 & 2.729 & 2.594 & 2.572 & 2.615 & 2.498 \\ 
		\hline
		Data  & Soft BART & GP & GP  \newline 5 vars & GP Tobit & Soft  TOBART & Soft TOBART  NP \\ 
		\hline
		\multicolumn{7}{c}{normal distribution, $\sigma=1$} \\
		\hline
		\cite{friedman1991multivariate} & 0.942 & 50.248 & 0.985 & 50.331 & \textbf{0.743} & 0.745 \\ 
		\cite{friedman1991multivariate} 1 side & 1.101 & 85.594 & 1.076 & 85.532 & 0.929 & \textbf{0.927} \\ 
		\cite{groot2012gaussian} & 2.187 & 16.045 & 2.728 & 16.042 & \textbf{0.546} & \textbf{0.546} \\ 
		\cite{jacobson2022high} & 0.684 & 7.109 & 0.681 & 7.053 & \textbf{0.675} & \textbf{0.675} \\ 
		\cite{sigrist2019grabit} & 1.002 & 2.460 & 0.949 & 2.487 & \textbf{0.973} & \textbf{0.973} \\ 
		\hline
		\multicolumn{7}{c}{Skew-t distribution, $location = 1$, $scale = 1$, $\nu=4$}\\
		\hline
		\cite{friedman1991multivariate} & 1.250 & 47.049 & 1.316 & 47.099 & 1.066 & \textbf{1.027} \\ 
		\cite{friedman1991multivariate} 1 side & 1.579 & 81.482 & 1.604 & 81.415 & 1.405 & \textbf{1.367} \\ 
		\cite{groot2012gaussian} & 2.769 & 19.210 & 3.348 & 19.210 & 0.947 & \textbf{0.933} \\ 
		\cite{jacobson2022high} & 1.170 & 8.877 & 1.138 & 8.823 & 1.142 & \textbf{1.133} \\ 
		\cite{sigrist2019grabit} & 1.146 & 4.487 & 1.100 & 4.556 & 1.135 & \textbf{1.127} \\ 
		\hline
		\multicolumn{7}{c}{Weibull distribution, $shape = 0.5$, $scale = 0.2$} \\
		\hline
		\cite{friedman1991multivariate} & 0.721 & 46.692 & 0.758 & 46.730 & 0.492 & \textbf{0.436} \\ 
		\cite{friedman1991multivariate} 1 side & 1.054 & 80.639 & 1.001 & 80.573 & 0.787 & \textbf{0.733} \\ 
		\cite{groot2012gaussian} & 1.712 & 15.435 & 1.903 & 15.436 & 0.610 & \textbf{0.598} \\ 
		\cite{jacobson2022high} & 0.718 & 7.447 & 0.716 & 7.392 & 0.705 & \textbf{0.691} \\ 
		\cite{sigrist2019grabit} & 0.338 & 2.708 & 0.338 & 2.744 & 0.340 & \textbf{0.305} \\ 
		\hline
		% \multicolumn{7}{c}{t distribution, $\nu = 3$} \\
		% \hline
		% \cite{friedman1991multivariate} & 2.016 & 53.098 & 2.153 & 53.172 & 1.880 & 1.864 \\ 
		% \cite{friedman1991multivariate} 1 side & 3.149 & 88.297 & 3.272 & 88.236 & 2.992 & 2.949 \\ 
		% \cite{groot2012gaussian} & 3.090 & 16.950 & 4.805 & 16.948 & 1.770 & 1.762 \\ 
		% \cite{jacobson2022high} & 2.311 & 10.274 & 2.234 & 10.218 & 2.261 & 2.229 \\ 
		% \cite{sigrist2019grabit} & 2.473 & 3.616 & 2.322 & 3.649 & 2.432 & 2.384 \\ 
		\hline
	\end{tabular}
	\caption{Simulation Study, Mean Squared Error. Minimum values, excluding GP trained with only the 5 relevant variables, are in bold.}
	\label{allsim_mse_tab}
\end{table*}

\begin{table*}[ht]
	\centering
	\begin{tabular}{l|p{1.1cm}p{1.1cm}p{1.5cm}p{1.1cm}p{1.3cm}p{2.2cm}}
		\hline
		Data  & Tobit & BART & RF & Grabit & TOBART & TOBART NP \\ 
		\hline
		\multicolumn{7}{c}{normal distribution, $\sigma=1$} \\
		\hline
		\cite{friedman1991multivariate} & 0.140 & 0.165 & 0.195 & 0.135 & 0.069 & 0.070 \\ 
		\cite{friedman1991multivariate} 1 side & 0.061 & 0.058 & 0.077 & 0.067 & 0.032 & 0.033 \\ 
		\cite{groot2012gaussian} & 0.287 & 0.121 & 0.171 & 0.158 & 0.116 & 0.115 \\ 
		\cite{jacobson2022high} & 0.099 & 0.104 & 0.123 & 0.105 & 0.102 & 0.102 \\ 
		\cite{sigrist2019grabit} & 0.052 & 0.052 & 0.053 & 0.050 & 0.047 & 0.047 \\ 
		\hline
		\multicolumn{7}{c}{Skew-t distribution, $location = 1$, $scale = 1$, $\nu=4$}\\
		\hline
		\cite{friedman1991multivariate} & 0.152 & 0.173 & 0.197 & 0.143 & 0.083 & 0.081 \\ 
		\cite{friedman1991multivariate} 1 side & 0.074 & 0.071 & 0.094 & 0.070 & 0.047 & 0.044 \\ 
		\cite{groot2012gaussian} & 0.281 & 0.122 & 0.172 & 0.177 & 0.119 & 0.119 \\ 
		\cite{jacobson2022high} & 0.098 & 0.103 & 0.121 & 0.141 & 0.106 & 0.101 \\ 
		\cite{sigrist2019grabit} & 0.049 & 0.049 & 0.050 & 0.049 & 0.047 & 0.046 \\ 
		\hline
		\multicolumn{7}{c}{Weibull distribution, $shape = 0.5$, $scale = 0.2$} \\
		\hline
		\cite{friedman1991multivariate} & 0.137 & 0.163 & 0.197 & 0.170 & 0.058 & 0.053 \\ 
		\cite{friedman1991multivariate} 1 side & 0.060 & 0.059 & 0.081 & 0.061 & 0.028 & 0.024 \\ 
		\cite{groot2012gaussian} & 0.290 & 0.058 & 0.144 & 0.116 & 0.077 & 0.073 \\ 
		\cite{jacobson2022high} & 0.056 & 0.061 & 0.088 & 0.072 & 0.072 & 0.046 \\ 
		\cite{sigrist2019grabit} & 0.046 & 0.045 & 0.046 & 0.040 & 0.041 & 0.038 \\ 
		\hline
		%     		 \multicolumn{7}{c}{t distribution, $\nu = 3$} \\
		% \hline
		% \cite{friedman1991multivariate} & 0.150 & 0.177 & 0.202 & 0.266 & 0.096 & 0.094 \\ 
		%   \cite{friedman1991multivariate} 1 side & 0.067 & 0.065 & 0.085 & 0.105 & 0.045 & 0.043 \\ 
		%   \cite{groot2012gaussian} & 0.284 & 0.145 & 0.186 & 0.202 & 0.143 & 0.139 \\ 
		%   \cite{jacobson2022high} & 0.120 & 0.123 & 0.136 & 0.140 & 0.124 & 0.120 \\ 
		%   \cite{sigrist2019grabit} & 0.054 & 0.054 & 0.055 & 0.055 & 0.052 & 0.051 \\ 
		\hline
		Data  & Soft BART & GP & GP  \newline 5 vars & GP Tobit & Soft  TOBART & Soft TOBART  NP \\ 
		\hline
		\multicolumn{7}{c}{normal distribution, $\sigma=1$} \\
		\hline
		\cite{friedman1991multivariate} & 0.116 & 0.477 & 0.120 & 0.477 & \textbf{0.055} & \textbf{0.055} \\ 
		\cite{friedman1991multivariate} 1 side & 0.033 & 0.455 & 0.055 & 0.456 & \textbf{0.025} & \textbf{0.025} \\ 
		\cite{groot2012gaussian} & 0.107 & 0.241 & 0.233 & 0.241 & \textbf{0.106} & 0.107 \\ 
		\cite{jacobson2022high} & 0.100 & 0.440 & 0.168 & 0.444 & \textbf{0.099} & \textbf{0.099} \\ 
		\cite{sigrist2019grabit} & 0.050 & 0.054 & 0.050 & 0.055 & \textbf{0.042} & \textbf{0.042} \\ 
		\hline
		\multicolumn{7}{c}{Skew-t distribution, $location = 1$, $scale = 1$, $\nu=4$}\\
		\hline
		\cite{friedman1991multivariate} & 0.135 & 0.489 & 0.144 & 0.489 & 0.064 & \textbf{0.062} \\ 
		\cite{friedman1991multivariate} 1 side & 0.047 & 0.459 & 0.078 & 0.459 & 0.034 & \textbf{0.033} \\ 
		\cite{groot2012gaussian} & 0.110 & 0.251 & 0.258 & 0.251 & \textbf{0.111} & \textbf{0.111} \\ 
		\cite{jacobson2022high} & 0.098 & 0.444 & 0.171 & 0.447 & 0.098 & \textbf{0.097} \\ 
		\cite{sigrist2019grabit} & 0.048 & 0.049 & 0.046 & 0.050 & \textbf{0.044} & \textbf{0.044} \\ 
		\hline
		\multicolumn{7}{c}{Weibull distribution, $shape = 0.5$, $scale = 0.2$} \\
		\hline
		\cite{friedman1991multivariate} & 0.120 & 0.488 & 0.120 & 0.488 & 0.037 & \textbf{0.032} \\ 
		\cite{friedman1991multivariate} 1 side & 0.031 & 0.463 & 0.060 & 0.463 & 0.016 & \textbf{0.012} \\ 
		\cite{groot2012gaussian} & 0.050 & 0.247 & 0.206 & 0.247 & 0.067 & \textbf{0.065} \\ 
		\cite{jacobson2022high} & 0.049 & 0.440 & 0.144 & 0.444 & 0.054 & \textbf{0.039} \\ 
		\cite{sigrist2019grabit} & 0.044 & 0.045 & 0.037 & 0.046 & 0.036 & \textbf{0.034} \\ 
		\hline
		% \multicolumn{7}{c}{t distribution, $\nu = 3$} \\
		% \hline
		% \cite{friedman1991multivariate} & 0.136 & 0.475 & 0.146 & 0.475 & 0.082 & 0.081 \\ 
		%   \cite{friedman1991multivariate} 1 side & 0.047 & 0.449 & 0.063 & 0.450 & 0.038 & 0.037 \\ 
		%   \cite{groot2012gaussian} & 0.133 & 0.246 & 0.329 & 0.246 & 0.133 & 0.132 \\ 
		%   \cite{jacobson2022high} & 0.120 & 0.431 & 0.182 & 0.434 & 0.118 & 0.116 \\ 
		%   \cite{sigrist2019grabit} & 0.053 & 0.055 & 0.054 & 0.056 & 0.050 & 0.050 \\ 
		\hline
	\end{tabular}
	\caption{Simulation Study, Brier Score. Minimum values, excluding GP trained with only the 5 relevant variables, are in bold.}
	\label{allsim_brier_tab}
\end{table*}

\subsection{Description of Treatment Effect Simulations}

A number of recent simulation studies have demonstrated that BART is among the most accurate treatment effect estimation methods \citep{wendling2018comparing, mcconnell2019estimating, dorie2019automated,hahn2019atlantic}. However, in practice many data sets, including randomized trial data sets, contain censored outcomes. For example, antibody concentrations or environmental levels of chemicals can only be measured accurately within a certain range as a result of limitations of measuring equipment. Often economic data is censored due to privacy considerations, for example income might be censored above a certain threshold. TOBART provides a machine learning treatment effect estimation method with uncertainty quantification that can be applied to this data while still making use of the information provided by censored observations. We demonstrate the effectiveness of TOBART by censoring the outcomes of DGPs from published studies of machine learning methods for treatment effect estimation. The chosen data generating processes contain linear and non-linear functions of covariates, constant and heterogeneous effects, and various degrees of confounding.

\subsubsection{ Censored  \cite{caron2022shrinkage} Simulations }

$P=10$ covariates are generated from a multivariate Gaussian distribution, $X_1,\hdots, X_{10} \sim \mathcal{MVN}(\bm{0}, \Sigma)$, with $\Sigma_{jk} = 0.6^{|j-k|} + 0.1 \mathbb{I}(j \neq k) $. The binary treatment variable is Bernoulli distributed, $Z_i \sim \text{Bern}(\pi(\bm{x}_i)) $, where
$$ \pi(\bm{x}_i) = \Phi(-0.4 + 0.3 X_{i,1} + 0.2 X_{i,2} ) $$
and $\Phi(\cdot)$ is the cumulative distribution function of the standard normal distribution.

\noindent The prognostic score function, $\mu(\bm{x}_i)$, and CATE function, $\tau(\bm{x}_i)$, are defined as
$$ \mu(\bm{x}_i) = 3 + X_{i,1} + 0.8 \sin ( X_{i,2} ) + 0.7 X_{i,3} X_{i,4} - X_{i,5}$$
$$ \tau(\bm{x}_i) = 2 + 0.8 X_{i,1} - 0.3 X_{i,2}^2 $$
The outcome before censoring is generated as:
$$ Y_i^* = \mu(\bm{x}_i) + \tau(\bm{x}_i)  Z_i  + \varepsilon_i \ , \ \text{where} \ \varepsilon_i \sim \mathcal{N}(0,1) $$
The number of sampled observations is 200. The observed outcome $Y_i$ is censored from below at the $15^{th}$ percentile of the generated $Y_i^*$ values, and from above at the $85^{th}$ percentile.

\subsubsection{Censored \cite{friedberg2020local} Simulations}

$P=20$ covariates are generated from independent standard uniform distributions $X_1,...,X_{20} \sim \mathcal{U}[0,1]$. There is no confounding as $\pi(\bm{x}_i)=0.5$ and $Z_i \sim \text{Bern}(\pi(\bm{x}_i)) $.     
\noindent The prognostic score function, $\mu(\bm{x}_i)$, and CATE function, $\tau(\bm{x}_i)$, are defined as $ \mu(\bm{x}_i) = 0  $ and
$$ \tau(\bm{x}_i) = \Bigg( 1 + \frac{1}{1 + \exp \Big(-20(X_{i,1} - \frac{1}{3}) \Big)} \Bigg) \times $$
$$\Bigg( 1 + \frac{1}{1 + \exp \Big(-20(X_{i,2} - \frac{1}{3}) \Big)} \Bigg).  $$ 
The outcome before censoring is generated as:
$$ Y_i^* = \mu(\bm{x}_i) + \tau(\bm{x}_i)  Z_i  + \varepsilon_i \ , \ \text{where} \ \varepsilon_i \sim \mathcal{N}(0,1) $$
The number of sampled observations is 200. The observed outcome $Y_i$ is censored from below at the $15^{th}$ percentile of the generated $Y_i^*$ values, and from above at the $85^{th}$ percentile.

\subsubsection{Censored \cite{nie2021quasi} Simulations}

The covariates are generated as follows across scenarios A to D. In simulation A, $X_1,...,X_{12} \sim \mathcal{U}[0,1]$. In simulations B to D, $X_1,...,X_{12} \sim \mathcal{N}(0,1)$.
%
% \begin{itemize}
%     \item[A.] $X_1,...,X_{12} \sim \mathcal{U}[0,1]$
%     \item[B.] $X_1,...,X_{12} \sim \mathcal{N}(0,1)$
%     \item[C.] $X_1,...,X_{12} \sim \mathcal{N}(0,1)$
%     \item[D.] $X_1,...,X_{12} \sim \mathcal{N}(0,1)$
% \end{itemize} 
%

$\pi(\bm{x}_i)$ is defined as follows across scenarios A to D: (A) $ \text{trim}_{0.1} \{ \sin(\pi X_{i,1} X_{i,2} )  \} $, (B) constant equal to $0.5$, (C) $1/\{1 + \exp(X_{i,2} + X_{i,3} )\}$, (D) $1/\{1 + \exp(-X_{i,1}) + \exp( - X_{i,2} )\}$.

% \medskip

% \noindent $\pi(\bm{x}_i)$ is defined as follows across scenarios A to D
% %
% \begin{itemize}
%     \item[A.] $ \text{trim}_{0.1} \{ \sin(\pi X_{i,1} X_{i,2} )  \} $
%     \item[B.] Constant equal to $0.5$
%     \item[C.] $\frac{1}{1 + \exp(X_{i,2} + X_{i,3} )}$
%     \item[D.] $\frac{1}{1 + \exp(-X_{i,1}) + \exp( - X_{i,2} )}$
% \end{itemize}
% %
% %

$\mu(\bm{x}_i)$ is defined as follows across scenarios A to D: (A) $\sin(\pi X_{i,1}  X_{i,2}) + 2 (X_{i,3}-0.5)^2 + X_{i,4} + 0.5 X_{i,5} $, (B) $\max \{X_{i,1} + X_{i,2}, X_{i,3} ,0   \} $, (C)  $2 \log \{ 1 + \exp ( X_{i,1}  + X_{i,2}  + X_{i,3} )   \} $, (D)  $\frac{1}{2} [ \max \{  X_{i,1}  + X_{i,2}  + X_{i,3}  ,0 \} + \max \{  X_{i,4}  + X_{i,5}  ,0 \} ]  $.

% \medskip

% \noindent $\mu(\bm{x}_i)$ is defined as follows across scenarios A to D
% %
% \begin{itemize}
%     \item[A.] $\sin(\pi X_{i,1}  X_{i,2}) + 2 (X_{i,3}-0.5)^2 + X_{i,4} + 0.5 X_{i,5} $
%     \item[B.] $\max \{X_{i,1} + X_{i,2}, X_{i,3} ,0   \} $
%     \item[C.] $2 \log \{ 1 + \exp ( X_{i,1}  + X_{i,2}  + X_{i,3} )   \} $
%     \item[D.] $\frac{1}{2} [ \max \{  X_{i,1}  + X_{i,2}  + X_{i,3}  ,0 \} + \max \{  X_{i,4}  + X_{i,5}  ,0 \} ]  $
% \end{itemize}
% %
% %

$\tau(\bm{x}_i)$ is defined as follows across scenarios A to D: (A) $ ( X_{i,1} +  X_{i,2})/2$, (B) $  X_{i,1} + \log \{1 + \exp( X_{i,2})  \}$, (C) constant equal to $ 1 $ , (D) $\max \{  X_{i,1}  + X_{i,2}  + X_{i,3}  ,0 \} - \max \{  X_{i,4}  + X_{i,5}  ,0 \}$.

% \medskip

% \noindent $\tau(\bm{x}_i)$ is defined as follows across scenarios A to D
% %
% \begin{itemize}
%     \item[A.] $ \frac{ X_{i,1} +  X_{i,2}}{2}$
%     \item[B.] $  X_{i,1} + \log \{1 + \exp( X_{i,2})  \}$
%     \item[C.] Constant equal to $ 1 $
%     \item[D.] $\max \{  X_{i,1}  + X_{i,2}  + X_{i,3}  ,0 \} - \max \{  X_{i,4}  + X_{i,5}  ,0 \}$
% \end{itemize}
% %
% %

% \medskip

\noindent The outcome before censoring is generated as:
$$ Y_i^* = \mu(\bm{x}_i) + \tau(\bm{x}_i)  (Z_i-0.5)  + \varepsilon_i \ , \ \text{where} \ \varepsilon_i \sim \mathcal{N}(0,1) $$
The number of sampled observations is 200. The observed outcome $Y_i$ is censored from below at the $15^{th}$ percentile of the generated $Y_i^*$ values, and from above at the $85^{th}$ percentile.

\subsection{Treatment Effect Simulation Results}

All methods are evaluated in terms of Precision in Estimation of Heterogeneous Effects (PEHE), which is defined as $\frac{1}{N}\sum_{i=1}^N (\hat{\tau}(\bm{x}_i) - \tau(\bm{x}_i) )^2$ . Confidence intervals are evaluated in terms of average coverage of $95\%$ intervals and average length of intervals.

The results are presented in Table \ref{causalsim_CaronFriedberg_tab}. For all DGPs, at least one TOBART method attains lower PEHE than all other methods, often by a large margin. Local Linear Forests \citep{friedberg2020local} attain similar PEHE to TOBART and TOBART-NP for \cite{nie2021quasi} DGP D, which involves partly linear prognostic and treatment effect functions, although soft TOBART is notably more accurate. The average coverages of TOBART and soft TOBART credible intervals for $\tau(\bm{x}_i)$ are generally much closer to $95\%$ than the coverages of intervals produced by competing methods. TOBART-NP produces very wide credible intervals relative to TOBART. TOBART-NP produces better coverage than TOBART for four DGPs.

% \subsubsection{Caron and LLF Simulations}

% latex table generated in R 4.1.0 by xtable 1.8-4 package
% Wed Oct 12 12:49:26 2022
\begin{table*}[t]
	\centering
	\begin{tabular}{|l|p{1cm}p{1cm}p{1cm}|p{1cm}p{1cm}p{1cm}|p{1cm}p{1cm}p{1cm}|}
		\hline
		& \multicolumn{3}{|p{4cm}|}{\cite{caron2022shrinkage} DGP} & \multicolumn{3}{|p{4cm}|}{\cite{friedberg2020local} DGP} &  \multicolumn{3}{p{4cm}|}{\cite{nie2021quasi} DGP A} \\ 
		\hline
		Method & PEHE & Cov. & Len. & PEHE & Cov. & Len.  & PEHE & Cov. & Len.  \\ 
		\hline
		BART all &  0.943 & 0.311 & 0.873 & 1.301 & 0.378 & 0.814 & 0.058 & 0.638 & 0.453   \\ 
		BART uncens. & 1.345 & 0.235 & 0.931  & 2.772 & 0.122 & 0.770 & 0.122 & 0.385 & 0.436 \\ 
		Soft BART all &  0.609 & 0.640 & 1.491  & 0.538 & 0.558 & 1.325 & 0.060 & 0.778 & 0.616  \\ 
		Soft BART uncens. &  1.048 & 0.428 & 1.412   & 2.067 & 0.294 & 1.139  & 0.145 & 0.387 & 0.445 \\  
		CF & 0.863 & 0.415 & 0.891 & 0.953 & 0.365 & 0.784  & 0.259 & 0.767 & 0.683  \\ 
		LLF &  0.850 & 0.462 & 1.069  & 0.876 & 0.443 & 1.037 & 0.281 & 0.791 & 0.776 \\ 
		Grabit & 0.722 & - & -  & 0.512 & - & - & 0.128 & - & -  \\ 
		TOBART & 0.536 & 0.863 & 1.917  & 0.329 & 0.900 & 1.905  & \textbf{0.043} & 0.993 & 1.121 \\ 
		TOBART-NP & 0.528 & \textbf{0.935} & 2.470 &  0.314 & \textbf{0.970} & 2.318 &  \textbf{0.043} & 1 & 1.655   \\ 
		Soft TOBART &  \textbf{0.390} & \textbf{0.935} & 1.970   & 0.218 & 0.905 & 1.738  & \textbf{0.043} & \textbf{0.968} & 0.867 \\ 
		Soft TOBART-NP & 0.394 & 0.972 & 2.657 & \textbf{0.211} & 0.9855 & 2.261 & \textbf{0.043} & 1 & 1.471 \\ 
		\hline
		\hline
		& \multicolumn{3}{|p{4cm}|}{\cite{nie2021quasi} DGP B} & \multicolumn{3}{|p{4cm}|}{\cite{nie2021quasi} DGP C} &  \multicolumn{3}{p{4cm}|}{\cite{nie2021quasi} DGP D} \\ 
		\hline
		Method & PEHE & Cov. & Len. & PEHE & Cov. & Len.  & PEHE & Cov. & Len.  \\ 
		\hline
		BART all & 1.041 & 0.2912 & 0.661 & 0.126 & 0.283 & 0.560  & 1.628 & 0.360 & 0.502  \\ 
		BART uncens. & 1.435 & 0.1972 & 0.579  & 0.319 & 0.048 & 0.591  & 1.699 & 0.346 & 0.461 \\ 
		Soft BART all & 0.548 & 0.577 & 0.976 & 0.102 & 0.516 & 0.700 & 1.102 & 0.565 & 1.181 \\ 
		Soft BART uncens. & 0.966 & 0.398 & 0.929  & 0.265 & 0.291 & 0.804  & 1.674 & 0.340 & 0.423  \\ 
		CF  & 0.650 & 0.686 & 1.148 & 0.555 & 0.013 & 0.543  & 1.165 & 0.391 & 0.722   \\ 
		LLF  & 0.550 & 0.846 & 1.570  & 0.543 & 0.038 & 0.563  & 0.933 & 0.535 & 1.096 \\ 
		Grabit & 0.650 & - & - &  0.242 & - & -  & 1.500 & - & - \\
		TOBART & 0.305 & 0.902 & 1.694 & 0.032 & \textbf{0.994} & 1.192 & 0.948 & 0.759 & 1.913  \\ 
		TOBART-NP & 0.316 & \textbf{0.954} & 2.307  & 0.032 & 1 & 2.114 &  0.937 & 0.818 & 2.304  \\ 
		Soft TOBART & \textbf{0.145} & 0.968 & 1.480 & \textbf{0.025} & 0.996 & 0.975 & \textbf{0.679} & 0.807 & 1.897 \\ 
		Soft TOBART-NP & 0.153 & 0.986 & 2.231 & \textbf{0.025} & 1 & 1.975 & 0.684 & \textbf{0.864} & 2.252  \\ 
		\hline
		
	\end{tabular}
	\caption{Treatment Effect Simulation Results.  \cite{caron2022shrinkage}, \cite{friedberg2020local}, and \cite{nie2021quasi} simulations with censoring from below at $15^{th}$ percentile and from above at $85^{th}$ percentile. PEHE = Precision in Estimation of Heterogeneous Effects, Cov = average coverage of 95\% intervals, Len = average length of 95\% intervals.}
	\label{causalsim_CaronFriedberg_tab}
\end{table*}

\section{Data Application}\label{application_sec}

For the data application, we consider the same methods as in section \ref{sim_results}, excluding Gaussian Processes and adding a hurdle model combining linear regression and probit. For each data set, we average results over 10 training-test splits. Each split is defined by taking a random sample of $\text{floor}(0.7n)$ training observations stratified by censorship status. Categorical variables were encoded as sets of dummy variables. 
% Many of these data sets do not involve true censoring and may be better modeled by hurdle models, or in the case of sample selection, type II Tobit models. Nonetheless, these data sets are often used for the purpose of demonstrating the applicability of censored regression models. 
The numbers of observations, covariates, and proportions of censored observations are given in table \ref{data_summary_tab}. Appendix \ref{data_desc_app} contains descriptions of each data set with references to original sources.

\subsection{Data Application Results}

The data application results are presented in Table \ref{data_mse_tab_relative} . For most data sets the results are similar across methods, particularly when methods are evaluated in terms of Brier score for predicted probabilities of censoring. Similar results can be observed for the AUC in the Supplementary Appendix (Online Resource 1). TOBART can give notably lower MSE of outcome predictions relative to other methods for some data sets. 

In contrast to the simulation studies above, there is not a clear winning method in Table \ref{data_mse_tab_relative}. Although censored outcome models have been applied to these data sets in previous work, perhaps other models are more suitable for some data sets. This is evidenced by the fact that for many data sets the combination of probit and a linear model outperforms Tobit. Therefore for some data sets zero inflated, hurdle, or sample selection models might be more appropriate. For the data sets on which Tobit outperforms probit and a linear model in terms of MSE, namely \texttt{Recon} and \texttt{Atrazine}, the best method is Soft TOBART. The TOBART models also notably outperform other methods when applied to the \texttt{BostonHousing} and \texttt{Missouri} data sets.

A lesson from this study is that it is important to select the appropriate model for the data set. The TOBART and Grabit methods are designed for the same form of DGPs, therefore it is arguably fairer to compare these two methods. Soft TOBART produces lower MSE predictions than Grabit across almost all data sets.\footnote{Potentially Grabit could produce better results with more hyperparameter tuning, although this would be computationally costly.} Nonetheless, the results are less impressive than those observed in the simulation study. Possible explanations for this include slow mixing of the TOBART Markov Chain, small sample sizes for some data sets, and very small or very large proportions of censored outcomes.% limiting the amount of potential gain in accuracy from Tobit-based models, large proportions of censored outcomes limiting the ability of any model to learn the response surface.%, and the  fact that some of these data sets do not arise from true censoring. For some data sets, zero inflated, hurdle, or sample selection models might be more appropriate.

%A lesson from this simulation study is that it is important to select the appropriate model for the data set at hand. The TOBART and Grabit methods are designed for the same form of DGPs, so it is arguably fairer to compare these two methods. %TOBART can give large improvements in MSE relative to Grabit, while Grabit only gives predictions that are more accurate by a large margin for one data set, \texttt{Workinghours}.

\begin{table}[t]
	\centering
	% \begin{adjustwidth}{-1cm}{}
	\begin{tabular}{l|l|l|p{0.9cm}|p{0.9cm}}
		\hline
		Data Set & n & p & \% cens. \newline below  & \% cens. \newline above \\ 
		\hline
		\texttt{antibody} & 330  & 3  & 26.1 & 0  \\ 
		% \texttt{Mofa} & 43  & 3  & 27.9  & 0  \\ 
		% \texttt{Affairs} & 601  & 20  & 75.0 & 0 \\ 
		\texttt{Recon} & 423  & 108  &  11.1 & 0 \\ 
		\texttt{Atrazine}& 48  &  2  & 29.2 & 0  \\ 
		\texttt{SedPb} & 42  & 2  &  3.6 &  0 \\ 
		\texttt{Pollen\_Thia} & 204  & 4  & 42.6 & 0 \\ 
		\texttt{Missouri} & 127 & 3  &  26.8 & 0 \\ 
		\texttt{BostonHousing} & 506  & 108  & 0 & 3.2 \\ 
		% \texttt{aptitude} & 200  & 5  &   0 & 8.5  \\ 
		% \texttt{jtrain2} &  445  & 17  &  30.8 & 0 \\ 
		% \texttt{Mroz} & 753  & 13  & 43.2  & 0 \\ 
		% \texttt{jtrain98}  &  1130 &  9  & 17.2  & 0 \\ 
		\hline
	\end{tabular}
	\caption{Data Application: Number of observations ($n$), number of covariates ($p$), and proportions censored from below and above.% for all data sets.
	}
	\label{data_summary_tab}
	% \end{adjustwidth}
\end{table}

% latex table generated in R 4.1.0 by xtable 1.8-4 package
% Tue Oct 10 20:49:17 2023
\begin{table*}[t]
	\centering
	% \begin{adjustwidth}{-1cm}{}
	\begin{tabular}{l|l|l|p{0.9cm}p{0.9cm}p{0.9cm}p{0.9cm}p{0.9cm}p{0.85cm}p{0.7cm}p{0.9cm}p{1cm}}
		\hline
		Data Set & n & p & TO \newline BART & TO \newline BART \newline NP & Soft \newline  TO \newline BART & Soft \newline TO\newline BART \newline NP & Grabit & BART & RF & Probit \newline +LM & Tobit \\ 
		\hline
		\multicolumn{12}{c}{MSE Relative to TOBART} \\
		\hline
		\texttt{antibody} & 330  & 3  &  1.00 & 0.98 & 1.01 & 0.97 & \textbf{0.96} & 1.00 & 0.99 & 0.99 & 0.99 \\ 
		% \texttt{Mofa} & 43  & 3  & 1.00 & \textbf{0.99} & 1.69 & 1.76 & 1.77 & 1.83 & 1.10 & 1.17 & 1.04 \\ 
		% \texttt{Affairs} & 601  & 20  & 1.00 & 0.96 & 1.05 & 1.01 & 1.08 & 0.78 & 0.77 & \textbf{0.76} & 1.01 \\ 
		\texttt{Recon} & 423  & 108  & 1.00 & 0.74 & \textbf{0.63} & 0.77 & 0.66 & 0.74 & 0.68 & 0.85 & 0.81 \\ 
		\texttt{Atrazine}& 48  &  2  & 1.00 & 0.98 & \textbf{0.97} & 0.98 & 0.99 & 1.01 & 1.01 & 1.02 & 1.00 \\ 
		\texttt{SedPb} & 42  & 2  & 1.00 & 0.97 & 0.94 & 0.95 & 1.03 & \textbf{0.93} & 1.08 & 1.02 & 1.02 \\ 
		\texttt{Pollen\_Thia} & 204  & 4  &  1.00 & \textbf{0.99} & 1.00 & 1.00 & \textbf{0.99} & \textbf{0.99} & \textbf{0.99} & \textbf{0.99} & 1.00 \\ 
		\texttt{Missouri} & 127 & 3  &   1.00 & 0.66 & \textbf{0.58} & 0.73 & 0.88 & 0.76 & 0.77 & 0.93 & 0.93 \\ 
		\texttt{BostonHousing} & 506  & 108  & \textbf{1.00} & 1.49 & 1.02 & 1.06 & 1.38 & 1.18 & 1.31 & 113.37 & 125.10 \\ 
		% \texttt{aptitude} & 200  & 5  &   1.00 & 1.00 & 0.99 & 0.99 & 1.12 & 0.97 & 1.06 & \textbf{0.89} & \textbf{0.89} \\ 
		% \texttt{jtrain2} &  445  & 17  &   1.00 & 0.97 & \textbf{0.95} & \textbf{0.95} & 1.14 & 0.97 & 1.02 & \textbf{0.95} & 1.00 \\ 
		% \texttt{Mroz} & 753  & 13  &   1.00 & \textbf{0.91} & 1.02 & \textbf{0.91} & 1.34 & 0.96 & 1.03 & 0.99 & 1.06 \\ 
		% \texttt{jtrain98}  &  1130 &  9  & 1.00 & 1.00 & \textbf{0.99} & \textbf{0.99} & 1.10 & \textbf{0.99} & 1.07 & \textbf{0.99} & 1.00 \\ 
		\hline
		\multicolumn{12}{c}{Brier Score} \\
		\hline
		\texttt{antibody} & 330  & 3  & 0.22 & 0.20 & 0.22 & 0.20 & 0.23 & 0.19 & 0.19 & 0.19 & 0.22 \\ 
		% \texttt{Mofa} & 43  & 3  & 0.20 & 0.21 & 0.20 & 0.25 & 0.23 & 0.13 & \textbf{0.11} & 0.17 & 0.21 \\ 
		% \texttt{Affairs} & 601  & 20  & \textbf{0.17} & \textbf{0.17} & \textbf{0.17} & \textbf{0.17} & 0.18 & \textbf{0.17} & 0.18 & 0.18 & \textbf{0.17} \\ 
		\texttt{Recon} & 423  & 108  & 0.17 & 0.12 & 0.17 & 0.11 & 0.15 & \textbf{0.09} & \textbf{0.09} & 0.22 & 0.24 \\ 
		\texttt{Atrazine}& 48  &  2  & 0.11 & 0.12 & 0.19 & 0.17 & 0.22 & 0.01 & 0.02 & \textbf{0.00} & 0.10 \\ 
		\texttt{SedPb} & 42  & 2  & \textbf{0.04} & \textbf{0.04} & 0.05 & 0.05 & 0.07 & \textbf{0.04} & 0.05 & 0.05 & 0.05 \\ 
		\texttt{Pollen\_Thia} & 204  & 4  & 0.17 & 0.16 & 0.18 & 0.16 & 0.22 & \textbf{0.13} & \textbf{0.13} & \textbf{0.13} & 0.17 \\ 
		\texttt{Missouri} & 127 & 3  &   0.17 & 0.15 & 0.17 & 0.14 & 0.18 & \textbf{0.13} & \textbf{0.13} & 0.18 & 0.22 \\ 
		\texttt{BostonHousing} & 506  & 108  & 0.02 & 0.03 & \textbf{0.01} & 0.02 & 0.02 & 0.03 & 0.02 & 0.13 & 0.03 \\ 
		% \texttt{aptitude} & 200  & 5  &  \textbf{0.07} & \textbf{0.07} & \textbf{0.07} & \textbf{0.07} & \textbf{0.07} & \textbf{0.07} & 0.08 & \textbf{0.07} & 0.10 \\ 
		% \texttt{jtrain2} &  445  & 17  &  0.01 & \textbf{0.00} & 0.01 & \textbf{0.00} & 0.05 & \textbf{0.00} & 0.01 & \textbf{0.00} & 0.01 \\ 
		% \texttt{Mroz} & 753  & 13  &  0.20 & \textbf{0.19} & \textbf{0.19} & \textbf{0.19} & 0.22 & \textbf{0.19} & 0.20 & \textbf{0.19} & \textbf{0.19} \\ 
		% \texttt{jtrain98}  &  1130 &  9  &  0.01 & \textbf{0.00} & \textbf{0.00} & \textbf{0.00} & 0.02 & \textbf{0.00} & \textbf{0.00} & \textbf{0.00} & \textbf{0.00} \\ 
		\hline
	\end{tabular}
	\caption{Data Application Results: Mean Squared Error of Outcome Predictions Relative to TOBART, and Brier Score for Predicted Probabilities of Censoring. Average over 10 random splits into 70\% training data 30\% test data. The numbers of observations and covariates are denoted by $n$ and $p$ respectively.
	}
	\label{data_mse_tab_relative}
	% \end{adjustwidth}
\end{table*}

\section{Conclusion}\label{conclusion_sec}
%
%Type I TOBART improves on existing censored outcome methods in terms of predictive probabilities of censoring, prediction of outcomes, and treatment effect estimation. The fully nonparametric extension, TOBART-NP, gives better uncertainty quantification in some simulations. Advantages of TOBART over competing methods such as Grabit \citep{sigrist2019grabit} include the lack of hyperparameter tuning and uncertainty quantification. TOBART can be combined with other variations on BART to allow for smooth data generating processes \citep{linero2018bayesianA}, sparsity \citep{linero2018bayesianB}, or multiple outcomes \citep{chakraborty2016bayesian}.
%
%
Type I TOBART produces accurate predictive probabilities of censoring, predictions of outcomes, and treatment effect estimates. TOBART-NP, gives better uncertainty quantification for some simulated DGPs. Advantages of TOBART over competing methods include the fact that hyperparameter tuning is not required, and the straightforward combination of the method with other variations on BART to allow for smooth DGPs and sparsity \citep{linero2018bayesianB}.%, or multiple outcomes \citep{chakraborty2016bayesian}.

\backmatter

\bmhead{Supplementary information}

The online supplementary appendix contains (A) additional simulation study results, (B) additional data application results, and (C) implementation details and parameter settings.

% If your article has accompanying supplementary file/s please state so here. 

% Authors reporting data from electrophoretic gels and blots should supply the full unprocessed scans for key as part of their Supplementary information. This may be requested by the editorial team/s if it is missing.

% Please refer to Journal-level guidance for any specific requirements.

\bmhead{Acknowledgments}

The author gratefully acknowledges helpful comments from Mikhail Zhelonkin, Chen Zhou, and participants at the Econometric Institute internal seminar.

\section*{Declarations}

\textbf{Conflict of interest}: The authors declare no competing interests.

%Some journals require declarations to be submitted in a standardised format. Please check the Instructions for Authors of the journal to which you are submitting to see if you need to complete this section. If yes, your manuscript must contain the following sections under the heading `Declarations':

% \begin{itemize}
% \item Conflict of interest/Competing interests (check journal-specific guidelines for which heading to use)
% \item Ethics approval 
% \item Consent to participate
% \item Consent for publication
% \item Availability of data and materials
% \item Authors' contributions
% \end{itemize}

% \noindent
% If any of the sections are not relevant to your manuscript, please include the heading and write `Not applicable' for that section. 

% \bigskip

% Springer journals and proceedings: \url{https://www.springer.com/gp/editorial-policies}

% Nature Portfolio journals: \url{https://www.nature.com/nature-research/editorial-policies}

% \textit{Scientific Reports}: \url{https://www.nature.com/srep/journal-policies/editorial-policies}

% BMC journals: \url{https://www.biomedcentral.com/getpublished/editorial-policies}
% \end{flushleft}

\FloatBarrier

\begin{appendices}
	
	\section{TOBART-1 Gibbs Sampler}\label{mcmc_app}
	
	\subsection{Gibbs Sampler Algorithms}
	
	For completeness of exposition, we describe here the full conditional samples from $p((T_k, M_k)| \{(T_j,M_j)\}_{j \neq k}, \sigma, \bm{y}^*) \ k = 1, \dots, m $ introduced by \cite{chipman2010bart} in Algorithm \ref{alg:treefullcond}. This sample is separated  into a Metropolis-Hastings draw of  $p(T_k| \{(T_j,M_j)\}_{j \neq k}, \sigma, \bm{y}^*) \ k = 1, \dots, m $ following by a closed form (multivariate normal) draw from \newline $p( M_k| T_k, \{(T_j,M_j)\}_{j \neq k}, \sigma, \bm{y}^*) \ k = 1, \dots, m)$. 
	
	The TOBART and TOBART-NP Gibbs samplers are described in algorithm \ref{alg:TOBART}.
	
	\begin{algorithm*}
		\caption{Full conditional sampler for Bayesian trees \citep{chipman2010bart}}\label{alg:treefullcond}
		\begin{algorithmic}
			\State \textbf{Input:} (Latent) outcome values $\bm{y}^*$, covariates $\bm{X}$, constant error variance for TOBART $\sigma^2$, trees and terminal node parameters $\{(T_k,M_k)\}_{k = 1}^m$  or observation-specific error means and variances for TOBART-NP $\{ \gamma_i, \sigma_i^2 \}_{i=1}^n$ . 
			\For{$k = 1,\dots m $}
			\State 1. Create partial residuals. For TOBART $ R_{k,i} = y_i^* - \sum_{s \neq k} g_k (x_i) $. For TOBART-NP $ R_{k,i} = y_i^* - \gamma_i - \sum_{s \neq k} g_k (x_i) $.
			\State 2. Draw from  $T_k | \{(T_j,M_j)\}_{j \neq k}, \sigma, \bm{y}^* $ using a Metropolis-Hastings Sampler. Propose $T_k'$ using a \texttt{PRUNE}, \texttt{CHANGE}, or \texttt{GROW} proposal. 
			\newline The \texttt{PRUNE} proposal removes (uniformly at random) a split that results in two terminal nodes from $T_k$. i.e. remove the children nodes from a random internal node without grandchildren.
			\newline The \texttt{CHANGE} proposal uniformly at random selects an internal node without grandchildren from tree $T_k$ and randomly samples a new splitting variable (uniformly) and splitting point (uniformly).
			\newline The \texttt{GROW} proposal uniformly at random selects a terminal node  of tree $T_k$ (with some minimum number of observations) and uniformly at random samples a new splitting variable and splitting point to create tree $T_k'$.
			
			\State 3. The log-likelihoods of trees $T_k$ and $T_k'$, marginalizing out the terminal node parameters $M_k$ (or $M_k'$), are standard weighted linear regression log-likelihoods $    \log \left(p(\mathbf{R}_k | T_k, M_k, \{\sigma_i^2\}_{i=1}^n)\right)   = $
			$$\sum_{l=1}^{b_k} \left[ - |n_{kl}| \log\left(  \sqrt{2\pi} \right) - \sum_{i\in n_{kl}} \log( \sigma_{i} )      - \frac{1}{2} \log\left( 1 +   \sum_{i\in n_{kl}} \frac{\sigma^2_\mu}{\sigma_{i}^2}  \right) - \sum_{i\in n_{kl}}  \frac{R_{ik}^2}{2 \sigma_{i}^2}       + \frac{  \left( \sum_{i \in n_{kl}} \frac{ R_{ik}}{\sigma_{i}^2} \right)^2 }{2\left(  \frac{1}{\sigma^2_\mu} +  \sum_{i \in n_{kl}} \frac{1}{\sigma_{i}^2}  \right)} \right] $$
			$b_k$ is the number of terminal nodes in $T_k$. There are $n_{kl}$ observations in the $l^{th}$ terminal node. \newline 
			\State 4. The Metropolis-Hastings step accepts the new tree proposal $T_k'$ with probability equal to $\frac{p(T_k' \rightarrow T_k)}{p(T_k \rightarrow T_k')}  \frac{ p(\mathbf{R}_k | T_k', M_k, \{\sigma_i^2\}_{i=1}^n) }{ p(\mathbf{R}_k | T_k, M_k, \{\sigma_i^2\}_{i=1}^n) }   \frac{ p(T_k') }{ p(T_k) } $, where $p(T_k \rightarrow T_k')$ is the probability of proposing tree $T_k'$ given the current tree is $T_k$. 
			\State  The terminal node parameters $M_k = (\mu_{1k},\dots,\mu_{b_k k})'$ are drawn from the full conditional  $p( M_k| T_k, \{(T_j,M_j)\}_{j \neq k}, \sigma, \bm{y}^*) \ k = 1, \dots, m)$, which separates into independent univariate normal draws. For $\ell = 1, \dots, b_{k}$ sample from
			$$  p(\mu_{\ell k} |  \{ R_{k,i} \}_{i \in \ell}, \{\sigma_i^2\}_{i=1}^n, \sigma_{\mu}^2  ) = \mathcal{N}(\mu_{\ell} | \tilde{\mu}_{\ell} ,  \tilde{\sigma}_{\ell}^2) \ , \   \tilde{\sigma}_{\ell}^2 =  \frac{1}{ \frac{1}{\sigma_{\mu}^2} +  \sum_{i \in \ell} \frac{1}{\sigma_i^2}  } , \tilde{\mu}_{\ell} =  \tilde{\sigma}_{\ell}^2 \Bigg(  \sum_{i \in \ell}  \frac{R_{k,i}}{\sigma_i^2}   \Bigg) $$

			\EndFor
			
		\end{algorithmic}
	\end{algorithm*}

	% \subsection{TOBART and TOBART-NP Gibbs Sampler}

	\begin{algorithm*}
		\caption{TOBART and TOBART-NP Gibbs Sampler}\label{alg:TOBART}
		\begin{algorithmic}
			\State \textbf{Input}: Number of MCMC iterations, $B$. Outcome values $\bm{y}$, covariates $\bm{X}$, censoring limits $a$ and $b$,  hyperparameter values $\alpha$, $\beta$, $\kappa$, $\nu$, $\lambda$ (or $q$). For TOBART-NP, hyperparameters $\gamma_0$ and $k_0$. 
			
			0. Set initial values of $f_y$, intialize all $m$ trees as stumps with no splits. For TOBART, initialize $\sigma$. For TOBART-NP, set initial values of $\gamma_i, \sigma_i $. % and start with all observations in a single cluster, with a draw from the prior $G_0$ for the initial parameter values.
			
			\For{$b = 1,\dots B $}
			\State 1. Draw latent variable from $\bm{y}^*| \gamma_i, \sigma_i, f_y, \bm{y} $
			$ y_i^* \sim 
			\begin{cases}
				\mathcal{TN}_{(-\infty, a]} \big(f(x_i)+\gamma_i, \sigma_i^2 \big)  \text{ if } y_i = a \\
				y_i  \text{ if }  a < y_i^* < b \\
				\mathcal{TN}_{(b, \infty]}\big(f(x_i)+\gamma_i, \sigma_i^2 \big)  \text{ if } y_i = b \\
			\end{cases} $
			\State 2. Draw the sum of trees $f_y | \bm{y}^*, \gamma_i, \sigma_i, $ by applying Algorithm \ref{alg:treefullcond}.
			\State 3. [TOBART] Draw $\sigma^2$ from an inverse gamma distribution $IG \Big(\frac{n + \nu}{2}, \frac{ \sum_{i=1}^n (y_i^* - \hat{y}_i)^2 + \nu \lambda}{2} \Big) $. 
			\State 3. [TOBART-NP] 
			\For{$i=1,...,n$} 
			\State Sample from $ (\gamma_i, \sigma_i) |  \bm{y}^*, \{ (\gamma_k, \sigma_k), k \neq i \}, f_y $. This follows the procedure described in \cite{escobar1994estimating, escobar1995bayesian, escobar1998computing}. For a similar context see step 3 of algorithm 1 of \cite{chib2010additive}.  Define $\vartheta_i = (\gamma_i, \sigma_i)$. Let $\vartheta_{-i} = \{(\gamma_{-i}, \sigma_{-i}) \} = \{ (\gamma_k, \sigma_k), k \neq i \}$ be the set of pairs of parameters excluding $(\gamma_i, \sigma_i)$. % \todo[]{Check if this also excludes elements for other individuals equal to $(\gamma_i, \sigma_i)$ }.  
			Let $(\gamma_{-i,r}^*, \sigma_{-i,r}^*)$ with $r = 1,..., k_{-i}$ be the set of $k_{-i}$ unique pairs of parameters in the set $\{(\gamma_{-i}, \sigma_{-i}) \}$.
			\begin{itemize}
				\item[(i)] Calculate $q_{i,0} = \alpha  t_{\nu}\Big(y_i^* | f_y (x_i), \lambda(1+\frac{1}{k_0})\Big)$ where $t_{\nu}$ denotes the probability density function of a t distribution with $\nu$ degrees of freedom.
				\item[(ii)] For $r=1,...,k_{-i}$ calculate $q_{-i,r} = n_{-i,r} \mathcal{N} \Big(y_i^* | \gamma_{-i,r}^* + f_y(x_i), (\sigma_{-i,r}^*)^2  \Big)  $.
				\item[(iii)] Scale $q_{i,0}$ and $q_{-i,r} $ to $ \tilde{q}_{i,0} = \frac{q_{i,0}}{q_{i,0} + \sum_{r=1}^{k_{-i}} q_{-i,r} }\ , \ \text{and} \ , \  \tilde{q}_{-i,r} = \frac{q_{-i,r} }{q_{i,0} + \sum_{s=1}^{k_{-i}} q_{-i,s} }  \ \text{for} \ r=1,...,k_{-i}$.
				\item[(iv)] Draw $r' \in \{ 0,1,...,k_{-i}\}$ from a categorical distribution with probabilities $ \{ \tilde{q}_{i,0}, \tilde{q}_{i,1}, ..., \tilde{q}_{i,k_{-i}}  \}$
				\item[(v)]
				$\begin{cases}
					\text{If } \ r' \neq 0, \ \text{ set } \ (\gamma_i, \sigma_i) = (\mu_{r'}, \sigma_{r'}) \\  
					\text{If } \ r' = 0, \text{ draw } \sigma_i^2 \sim \mathcal{IG} \Bigg( \frac{\nu +1}{2}, \frac{\nu \lambda}{2} + \frac{ (y_i^* - f_y (x_i))^2 }{ 2 \Big( 1 + \frac{1}{k_0} \Big) }  \Bigg) \text{ then  } \gamma_i | \sigma_i^2 \sim \mathcal{N} \Big( \frac{1}{k_0+1} (y_i^* - f_y (x_i)) , \frac{\sigma_i^2}{k_0+1}   \Big) \\
				\end{cases} $
				%\todo[inline]{Check if have to update the elements of $\{(\gamma_{-i}, \sigma_{-i}) \}$  after each draw or just after whole set of draws for all $i$. ANSWER: UPDATE AFTER DRAW EACH $i$. Maybe more computationally efficient to update counts of unique elements?}
			\end{itemize}
			\EndFor
			\State 4. [TOBART-NP] The following mixing step speeds up convergence of the Markov chain. Steps of this form were introduced by \cite{bush1996semiparametric} and \cite{west1994hierarchical}.
			
			Let $n_j$ denote the number of observations in cluster $j$, $N_j = \{i : \varrho_i = j\}$ where the variable $\varrho_i$ equals the index of the cluster to which observation $i$ belongs. Let $u_i = y_i^* - f_y(x_i)$ and let $\bar{u}^{(j)} = \frac{\sum_{i \in N_j} u_i}{n_j}$ be the mean of $u_i $ values for all observations in cluster $N_j$.
			
			Note that $ p(\gamma_j^*, \sigma_j^* | \bm{y}^*, f_y) \propto \prod_{i=1}^{n_j} \mathcal{N} \Big( y_i^* | f_y(x_i) + \gamma_j, \sigma_j^2 \Big) \mathcal{N} \Big( \gamma_j | 0, \frac{\sigma_j^2}{k_0} \Big) \mathcal{IG}\Big(\sigma_j^2 | \frac{\nu}{2}, \frac{\nu \lambda}{2} \Big) $	. 
			% or $ p(\gamma_j^*, \sigma_j^* | \bm{y}^*, f_y) \propto \prod_{i=1}^{n_j} \mathcal{N} \Big( u_i | \gamma_j, \sigma_j^2 \Big) \mathcal{N} \Big( \gamma_j | 0, \frac{\sigma_i^2}{k_0} \Big) \mathcal{IG}\Big(\sigma_j^2 | \frac{\nu}{2}, \frac{\nu \lambda}{2} \Big) $	
			A standard conjugacy result implies that we can sample
			$ {\sigma_j^*}^2 \sim \mathcal{IG}\Big( \frac{\nu + n_j}{2},  \frac{\nu \lambda}{2} + \frac{1}{2} \sum_{\forall i \in N_j} (u_i - \bar{u}^{(j)})^2 + \frac{n_j k_0}{k_0 + n_j} \frac{(\bar{u}^{(j)})^2}{2}   \Big) \ , \ \gamma_j^* \sim \mathcal{N} \Bigg( \frac{n_j \bar{u}^{(j)}}{k_0 +n_j}, \frac{ (\sigma_j^*)^2}{k_0 +n_j}  \Bigg) $
			\State 5. [TOBART-NP] Sample an auxiliary variable $\kappa \sim Beta(\alpha + 1 , n)$ and sample $\alpha$ from the mixture distribution
			$ \alpha | k \sim p_{\kappa} \text{Gamma} (c_1 +k , c_2 - \log \kappa ) + (1- p_{\kappa}) \text{Gamma}(c_1 +k -1 , c_2 - \log \kappa ) $
			where $k$ is the current number of mixture components, i.e. unique elements of $ \{ \vartheta_i \}_{i=1}^{n}=  \{ \gamma_i, \sigma_i \}_{i=1}^{n} $. $p_{\kappa}$ is the mixing probability, set so that
			$ \frac{p_{\kappa}}{1 -p_{\kappa} } = \frac{c_1 + k - 1}{n (c_2 - \log \kappa)} $. 
			%
			% giving
			% $ p_{\kappa} = \frac{ c_1 + k - 1  }{ n (c_2 - \log \kappa ) + c_1 + k - 1 } $
			%
			%
			If the prior on $\alpha$ is the prior applied by \cite{george2019fully, mcculloch2021causal} and \cite{conley2008semi}, then samples are obtained from $\alpha | k $ by noting that $p(\alpha | k) \propto p(k|\alpha) p(\alpha) \propto \alpha^k \frac{\Gamma(\alpha)}{\Gamma(n+\alpha)} \times (1- \frac{ \alpha - \alpha_{min}}{\alpha_{max} - \alpha_{min}} )^{\psi} $ \citep{antoniak1974mixtures}. A sample can be obtained by discretizing the support and making a multinomial draw. \cite{mcculloch2021causal} use an equally spaced grid of $100$ values from $\alpha_{min}$ to $\alpha_{max}$.
			
			\EndFor
			
		\end{algorithmic}
	\end{algorithm*}

	\subsection{TOBART-NP Out of sample distribution of the error}
	
	For test data predictive intervals, we may sample TOBART-NP error term values for out of sample observations. \cite{van2011bayesian} describes the sampling method as follows. Let $\tilde{n}$ denote the index of an out of sample observation. At iteration $t \in \{ 1,...,T\} $ of the Markov chain, given $\{\vartheta_{i,t}\}_{i=1}^n$, generate an out-of-sample value $\vartheta_{\tilde{n}, t}$ according to:
	$$ \vartheta_{\tilde{n}, t}  \begin{cases}
		= \vartheta_{i,t} \ \text{with probability} \ \frac{1}{\alpha + n} \ \text{for} \  i = 1,...,n \\
		\sim G_0  \ \text{with probability} \ \frac{\alpha}{\alpha + n}
	\end{cases} $$
	An estimate of the posterior predictive distribution of the error is 
	$$ \hat{f}(u |y, s) = \frac{1}{T} \sum_{t=1}^T f(u| \mu_{\tilde{n},t}, \sigma_{\tilde{n},t}^2) $$
	Also, samples $u_{\tilde{n}}^{(t)}$ can be made from $\mathcal{N} \Big( \mu_{\tilde{n},t}, \sigma_{\tilde{n},t}^2 \Big) $ for each iteration $t$ of the MCMC sampler.

	\section{Treatment Effect with Censored Outcomes; Additional Details}\label{TEbias_app}
	
	A model naively trained on the full dataset with censoring estimates the following:
	$$E[Y_i(1) - Y_i(0) |  \bm{x}_i] = \tau(\bm{x}_i) \times $$
	$$\Bigg(   \Phi \Big(\frac{b -  \mu(\bm{x}_i) - \tau(\bm{x}_i)  }{\sigma}  \Big) - \Phi \Big(\frac{a -  \mu(\bm{x}_i) - \tau(\bm{x}_i)      }{\sigma}  \Big) \Bigg)+ $$
	$$ \mu(\bm{x}_i) \Bigg(   \Phi \Big(\frac{b -  \mu(\bm{x}_i) - \tau(\bm{x}_i)  }{\sigma}  \Big) - $$
	$$ \Phi \Big(\frac{a -  \mu(\bm{x}_i) - \tau(\bm{x}_i)      }{\sigma}  \Big)  -    \Phi \Big(\frac{b - \mu(\bm{x}_i)}{\sigma}  \Big) $$
	$$ + \Phi \Big(\frac{a - \mu(\bm{x}_i)}{\sigma}  \Big)  \Bigg)   +$$
	$$ \sigma  \Bigg( \phi \Big(\frac{a - ( \mu(\bm{x}_i) + \tau(\bm{x}_i)))}{\sigma}  \Big) $$
	$$ - \phi \Big(\frac{b - ( \mu(\bm{x}_i) + \tau(\bm{x}_i))  }{\sigma}  \Big)   - 
	\phi \Big(\frac{a - \mu(\bm{x}_i)}{\sigma}  \Big) + $$
	$$ \phi \Big(\frac{b - \mu(\bm{x}_i)}{\sigma}  \Big) 
	\Bigg) $$
	$$ + a \Bigg(  \Phi \Big(\frac{a -  \mu(\bm{x}_i) - \tau(\bm{x}_i)  }{\sigma}  \Big) -  \Phi \Big(\frac{a -  \mu(\bm{x}_i)   }{\sigma}  \Big)  \Bigg) $$
	$$ + b \Bigg( \Phi \Big(\frac{b -  \mu(\bm{x}_i)   }{\sigma}  \Big) - \Phi \Big(\frac{b -  \mu(\bm{x}_i) - \tau(\bm{x}_i)  }{\sigma}  \Big) \Bigg) $$
	
	\FloatBarrier
	
	\FloatBarrier

	\clearpage
	
	\section{Description of Data Sets}\label{data_desc_app}
	
	\begin{itemize}%[label={}]
		% \item \texttt{aptitude}: This is a simulated data set obtained from \url{https://stats.oarc.ucla.edu/stata/dae/tobit-analysis/}. The outcome is an aptitude test score censored from above at 800, $n= 200$ and $p = 5$.
		% \item \texttt{Charity}: Charity donation data set available in the \texttt{R} package \texttt{wooldridge}, and sourced from \cite{franses2001quantitative}. The outcome is a charity donation bounded from below by $0$, $n= 4268$ and $p = 6$.
		% \item \texttt{jtrain2}: Training program data set available in the \texttt{R} package \texttt{wooldridge}, and sourced from \cite{lalonde1986evaluating}. The outcome is real earnings in 1978 bounded from below by $0$, $n= 445$ and $p = 17$.
		% \item \texttt{jtrain98}: Partly fake data available in the \texttt{R} package \texttt{wooldridge} similar to \texttt{jtrain2}. The outcome is real earnings in 1998 bounded from below by $0$, $n= 1130$ and $p = 9$.
		% \item \texttt{Mroz}: 1976 Panel Study of Income Dynamics data available in the \texttt{R} package \texttt{Ecdat} sourced from \cite{mroz1987sensitivity}. The outcome is  wife’s average hourly earnings in 1975 dollars, bounded from below by $0$, $n= 753$ and $p = 13$.
		% \item \texttt{RainIbk}: Weather data from SYNOP station Innsbruck Airport from 2000-01-01 to 2013-09-17 and forecasts from \cite{hamill2013noaa} available in the \texttt{R} package \texttt{crch}. The outcome is 3 days accumulated precipitation bounded from below by $0$, $n= 4971$ and $p = 13$. 
		\item \texttt{antibody}: Measles vaccine response data set obtained from \cite{moulton1995mixture}, originally from \cite{job1991successful}. The outcome is an antibody measurement censored from below at $0.1$, $n= 330$ and $p = 3$. 
		% \item \texttt{Mofa}: International Expansion of U.S. Majority–owned Foreign Affiliates in Fire (finance, Insurance and Real Estate) data available in the \texttt{R} package \texttt{Ecdat} sourced from \cite{ioannatos1995censored}. The outcome is capital expenditures, bounded from below by $0$, $n= 50$ and $p = 3$.
		% \item \texttt{Tobacco}: Tobacco consumption data available in the \texttt{R} package \texttt{Ecdat} sourced from \cite{verbeek2008guide}. The outcome is budget share of tobacco, bounded from below by $0$, $n= 2724$ and $p = 12$.
		% \item \texttt{Workinghours}: US 1987 labor supply data available in the \texttt{R} package \texttt{Ecdat} sourced from \cite{lee1995semi}. The outcome is wife's working hours per year, bounded from below by $0$, $n= 3382$ and $p = 14$.
		% \item \texttt{Affairs}: Psychology Today 1969 Extra-marital affairs survey data set available in the \texttt{R} package \texttt{AER} (also in \texttt{Ecdat} and \texttt{wooldridge})  sourced from \cite{fair1977note, fair1978theory}. The outcome is number of extramarital affairs in the past year, bounded from below by $0$, $n= 601$ and $p = 20$.\footnote{The continuous outcome models under consideration are not well suited to discrete outcome modelling. Nonetheless the \texttt{Affairs} data set is commonly used to illustrate continuous outcome Tobit models.}
		\item \texttt{Recon}: Atrazine concentrations in streams throughout the Midwestern United States. Data available in the \texttt{R} package \texttt{NADA} \citep{helsel2005nondetects} sourced from \cite{mueller1997logistic}. The outcome is Atrazine concentration, censored from below at $0.05$, $n= 423$ and $p = 108$.
		\item \texttt{Atrazine}: Atrazine concentrations in Nebraska ground water. Data available in the \texttt{R} package \texttt{NADA} \citep{helsel2005nondetects} sourced from \cite{junk1980areal}. The outcome is Atrazine concentration, censored from below at $0.01$, $n= 48$ and $p = 2$.
		\item \texttt{SedPb}: Lead concentrations in stream sediments before and after wildfires. Data available in the \texttt{R} package \texttt{NADA} \citep{helsel2005nondetects}. The outcome is Lead concentration, censored from below at $4$, $n= 82$ and $p = 2$.
		\item \texttt{Pollen\_Thia}: Thiamethoxam concentrations in pollen from the Ontario Pollen Monitoring Network. Data available in the \texttt{R} package \texttt{NADA2} \citep{helsel2005nondetects} sourced from \cite{junk1980areal}. The outcome is Thiamethoxam concentration, censored from below at $0.05$, $n= 204$ and $p = 4$.
		\item \texttt{Missouri}: TCDD concentrations used by \cite{zirschky1986geostatistical} in a geostatistical analysis of Hazardous waste data in Missouri. Data available in the \texttt{R} package \texttt{CensSpatial} \citep{helsel2005nondetects}. The outcome is censored from below at $0.1$, $n= 127$ and $p = 3$.
		% \item \texttt{medexp}: Medical Expenditure data provided with \cite{cameron2010microeconometrics}, sourced from \cite{deb2006bayesian}. Outcome is Ambulatory medical expenditures censored from below at $0$. $n = 3328 $, $p = 19$.
		\item \texttt{BostonHousing}: Housing data for 506 census tracts of Boston from the 1970 census available in the \texttt{R} package \texttt{mlbench} \citep{leisch2010machine}, sourced from \cite{harrison1978hedonic, pace1997using}. Outcome is median value of owner-occupied homes in USD 1000’s censored from above at $50$. $n = 506 $, $p = 108$.
	\end{itemize}

	\FloatBarrier
	
	\section{Comparison of Simulation Study Computational Times}\label{comp_time_app}
	
	This appendix contains a comparison of average computational times, in minutes, across iterations for each DGP of the simulation study. The times for BART, RF, and Soft BART do not contain the time taken to train separate models for estimation of binary censoring probabilities. The Grabit time does not contain the considerable time required for parameter tuning by 5-fold cross-validation (the model was re-trained 135 times in each fold for different parameter settings).
	
	The Gaussian Process \texttt{MATLAB} code written by \cite{groot2012gaussian} was called in \texttt{R} via the \texttt{R} package  \texttt{R.matlab}. All other functions were implemented in \texttt{R}. Therefore the Gaussian Process times are omitted for fair speed comparison. The Gaussian Process functions were fast, and ran for at most a few minutes per iteration.

	\begin{table*}[t]
		\centering
		\begin{tabular}{l|p{1.1cm}p{1.1cm}p{1.6cm}p{1.1cm}p{1.1cm}p{1.6cm}}
			\hline
			\hline
			Data  & Tobit & BART & RF & Grabit & Tobit \newline BART & Tobit \newline BART  NP \\ 
			\hline
			\cite{friedman1991multivariate} & 0.028 & 6.800 & 9.912 & 0.675 & 3.142 & 27.762 \\ 
			\cite{friedman1991multivariate} 1 side & 0.035 & 5.713 & 10.269 & 1.747 & 1.106 & 56.761 \\ 
			\cite{groot2012gaussian} & 0.167 & 6.046 & 40.916 & 1.153 & 33.681 & 26.193 \\ 
			\cite{jacobson2022high} & 0.030 & 7.508 & 13.267 & 0.364 & 3.563 & 11.993 \\ 
			\cite{sigrist2019grabit} & 0.030 & 6.262 & 11.670 & 0.265 & 13.910 & 41.027 \\ 
			\hline
			& Soft BART & Soft \newline Tobit \newline BART & Soft \newline Tobit \newline BART  NP \\ 
			\hline
			\cite{friedman1991multivariate} & 22.214 & 24.416 & 109.241 \\ 
			\cite{friedman1991multivariate} 1 side & 16.938 & 34.236 & 89.566 \\ 
			\cite{groot2012gaussian} & 30.325 & 89.028 & 103.573 \\ 
			\cite{jacobson2022high} & 33.373 & 9.905 & 150.921 \\ 
			\cite{sigrist2019grabit} & 17.918 & 44.855 & 97.271 \\ 
			\hline
		\end{tabular}
		\caption{Simulation Study, normal distribution, $\sigma=1$, Computational times, in minutes.}
		\label{normsim_sd1_time_tab}
	\end{table*}

	\begin{table*}[t]
		\centering
		\begin{tabular}{l|p{1.1cm}p{1.1cm}p{1.6cm}p{1.1cm}p{1.1cm}p{1.6cm}}
			\hline
			\hline
			& Tobit & BART & RF & Grabit & Tobit \newline BART & Tobit \newline BART NP \\ 
			\hline
			\cite{friedman1991multivariate} & 0.011  & 6.543 & 9.505 & 0.461 & 2.912 & 48.585 \\ 
			\cite{friedman1991multivariate} 1 side & 0.086 & 6.802 & 28.386 & 0.560 & 16.317 & 29.928 \\ 
			\cite{groot2012gaussian} & 0.026 & 6.438 & 11.567 & 0.352 & 3.527 & 20.167 \\ 
			\cite{jacobson2022high} & 0.054 & 6.813 & 19.126 & 0.404 & 8.625 & 19.915 \\ 
			\cite{sigrist2019grabit} & 0.068 & 7.536 & 22.446 & 0.353 & 10.291 & 79.636 \\ 
			\hline
			& Soft BART & Soft \newline Tobit \newline BART & Soft \newline Tobit \newline BART NP \\ 
			\hline
			\cite{friedman1991multivariate} & 18.040 & 40.479 & 88.207 \\ 
			\cite{friedman1991multivariate} 1 side & 32.417 & 31.089 & 117.085 \\ 
			\cite{groot2012gaussian} & 22.372 & 17.976 & 132.344 \\ 
			\cite{jacobson2022high} & 29.291 & 20.369 & 131.629 \\ 
			\cite{sigrist2019grabit} & 36.055 & 100.532 & 166.530 \\ 
			\hline
		\end{tabular}
		\caption{Simulation Study, Skew-t Distribution, Computational times, in minutes.}
		\label{sim_skewt_time_tab}
	\end{table*}

	\begin{table*}[t]
		\centering
		\begin{tabular}{l|p{1.1cm}p{1.1cm}p{1.6cm}p{1.1cm}p{1.1cm}p{1.6cm}}
			\hline
			\hline
			& Tobit & BART & RF & Grabit & Tobit \newline BART & Tobit \newline BART NP \\ 
			\hline
			\cite{friedman1991multivariate} & 0.011 & 6.705 & 17.705 & 0.489 & 9.170 & 39.305 \\ 
			\cite{friedman1991multivariate} 1 side & 0.094 & 7.075 & 30.974 & 0.634 & 21.699 & 21.012 \\ 
			\cite{groot2012gaussian} & 0.083 & 6.531 & 22.959 & 0.469 & 16.355 & 11.189 \\ 
			\cite{jacobson2022high} & 0.123 & 6.695 & 39.388 & 0.657 & 28.227 & 30.822 \\ 
			\cite{sigrist2019grabit} & 0.074 & 7.479 & 26.137 & 0.386 & 13.289 & 80.480 \\ 
			\hline
			& Soft BART & Soft \newline Tobit \newline BART & Soft \newline Tobit \newline BART NP \\ 
			\hline
			\cite{friedman1991multivariate} & 25.684 & 38.256 & 103.480 \\ 
			\cite{friedman1991multivariate} 1 side & 35.920 & 22.301 & 136.298 \\ 
			\cite{groot2012gaussian} & 33.716 & 10.948 & 130.280 \\ 
			\cite{jacobson2022high} & 33.999 & 12.964 & 120.808 \\ 
			\cite{sigrist2019grabit} & 39.141 & 12.418 & 151.876 \\ 
			\hline
		\end{tabular}
		\caption{Simulation Study, Weibull Distribution, Computational times, in minutes.}
		\label{sim_weibull_time_tab}
	\end{table*}

	% \FloatBarrier

	% \clearpage

	\section{Simulation Study - TOBART prior settings}\label{tbarts_sims_app}
	
	This appendix contains a comparison of simulation study results for different prior parameter settings.
	
	For standard TOBART, we present results for different $\lambda$ parameter settings. 
	Recall that $\sigma^{-2} \sim Ga(\frac{v}{2}, \frac{v \lambda}{2})$ and $\lambda$ is set such that the $q^{th}$ quantile of the prior distribution of $\sigma$ is equal to some estimate $\hat{\sigma}$. For standard BART, $\hat{\sigma}$ is the sample standard deviation of the residuals from a linear model. However, a standard linear model does not account for censoring, and therefore may give poor prior calibration.
	
	We consider the following options:
	
	\begin{itemize}
		\item naive sd: The sample standard deviation of the outcomes without accounting for censoring.
		\item Tobit sd: The maximum likelihood estimate of the standard deviation of the error term from a linear Tobit model (with covariates).
		\item cens sd: The maximum likelihood estimate of the standard deviation of the error term from an intercept-only linear Tobit model. This is an estimate of the standard deviation of $y^*$ that adjusts for censoring, assuming normality and no effects of covariates.
		\item lm sd: The default BART setting. The sample standard deviation of residuals from a linear model.
	\end{itemize}
	
	A limitation of the options that account for censoring is that the estimates rely on the assumption of normality. unsurprisingly, we observe that no setting provides the best results for all DGPs in table \ref{tbarts_prior_simstudy}. The $\hat{\sigma}$ estimate from an intercept-only Tobit model gives good results. It is generally larger than the estimates from other options and results in a less informative prior. Therefore we apply this option for our main results.
	
	\clearpage
	
	\begin{table}[ht]
		\centering
		\begin{tabular}{l|p{1.1cm}p{1.4cm}p{1.4cm}p{1.4cm}p{1.4cm}p{1.4cm}}
			\hline
			\hline
			Data  & Tobit \newline BART \newline naive \newline sd & Tobit \newline BART \newline Tobit \newline sd & Tobit \newline BART \newline cens \newline sd & Tobit \newline BART \newline lm \newline sd & Tobit \newline BART NP, \newline $\nu=3$, \newline Tobit sd & Tobit \newline BART NP, \newline $\nu=10$ \newline Tobit sd \\ 
			\hline
			\multicolumn{7}{c}{normal distribution, $\sigma=1$} \\
			\hline
			\cite{friedman1991multivariate} & 1.196 & 1.269 & 1.162 & 1.180 & 1.176 & 1.154 \\ 
			\cite{friedman1991multivariate} 1 side & 1.478 & 1.510 & 1.457 & 1.462 & 1.518 & 1.509 \\ 
			\cite{groot2012gaussian} & 0.638 & 0.634 & 0.631 & 0.644 & 0.625 & 0.617 \\ 
			\cite{jacobson2022high} & 0.717 & 0.717 & 0.718 & 0.716 & 0.720 & 0.720 \\ 
			\cite{sigrist2019grabit} & 1.072 & 1.072 & 1.072 & 1.075 & 1.075 & 1.074 \\ 
			\hline
			\hline
			\multicolumn{7}{c}{Skew-t distribution, $location = 1$, $scale = 1$, $\nu=4$}\\
			\hline
			\cite{friedman1991multivariate} & 1.715 & 1.725 & 1.648 & 1.696 & 1.677 & 1.597 \\ 
			\cite{friedman1991multivariate} 1 side & 2.220 & 2.307 & 2.163 & 2.200 & 2.079 & 2.060 \\ 
			\cite{groot2012gaussian} & 1.170 & 1.189 & 1.149 & 1.170 & 1.068 & 1.071 \\ 
			\cite{jacobson2022high} & 1.236 & 1.242 & 1.238 & 1.240 & 1.186 & 1.184 \\ 
			\cite{sigrist2019grabit} & 1.281 & 1.280 & 1.279 & 1.283 & 1.268 & 1.267 \\ 
			\hline
			\hline
			\multicolumn{7}{c}{Weibull distribution, $shape = 0.5$, $scale = 0.2$} \\
			\hline
			\cite{friedman1991multivariate} & 0.933 & 0.910 & 0.871 & 0.928 & 0.807 & 0.811 \\ 
			\cite{friedman1991multivariate} 1 side & 1.447 & 1.396 & 1.344 & 1.371 & 1.146 & 1.154 \\ 
			\cite{groot2012gaussian} & 0.771 & 0.782 & 0.787 & 0.846 & 0.649 & 0.648 \\ 
			\cite{jacobson2022high} & 0.942 & 0.920 & 0.960 & 0.906 & 0.697 & 0.696 \\ 
			\cite{sigrist2019grabit} & 0.417 & 0.424 & 0.426 & 0.424 & 0.365 & 0.362 \\ 
			\hline
			\hline
			\multicolumn{7}{c}{t distribution, $\nu = 3$} \\
			\hline
			\cite{friedman1991multivariate} & 2.406 & 2.498 & 2.459 & 2.420 & 2.339 & 2.369 \\ 
			\cite{friedman1991multivariate} 1 side & 3.761 & 3.786 & 3.667 & 3.679 & 3.629 & 3.538 \\ 
			\cite{groot2012gaussian} & 2.073 & 2.113 & 2.028 & 2.105 & 1.963 & 1.964 \\ 
			\cite{jacobson2022high} & 2.427 & 2.401 & 2.394 & 2.454 & 2.301 & 2.306 \\ 
			\cite{sigrist2019grabit} & 2.635 & 2.603 & 2.615 & 2.599 & 2.493 & 2.498 \\ 
		\end{tabular}
		\caption{Simulation Study, Mean Squared Error. Different Prior calibration settings for error term distribution.}
		\label{tbarts_prior_simstudy}
	\end{table}
	
	\FloatBarrier

	\clearpage
	% \section{Section title of first appendix}\label{secA1}
	
	% An appendix contains supplementary information that is not an essential part of the text itself but which may be helpful in providing a more comprehensive understanding of the research problem or it is information that is too cumbersome to be included in the body of the paper.
	
	
	
	
\end{appendices}


\FloatBarrier

\bibliography{main}% common bib file

\end{document}